\documentclass[journal]{IEEEtran}
\usepackage{cite}
\usepackage{threeparttable}
\usepackage{graphicx}
\usepackage{picinpar}
\usepackage[cmex10]{amsmath}
\usepackage{amsmath,amsfonts,amssymb}
\usepackage{algorithmic}
\usepackage{subfig}
\usepackage{algorithm,setspace}
\usepackage{threeparttable}
\usepackage{caption}
\usepackage{booktabs}
\usepackage{multirow}
\usepackage{diagbox}
\usepackage{etoolbox}
\usepackage{makecell}
\usepackage{amssymb}
\usepackage{mathrsfs}
\usepackage{color}
\definecolor{purple}{RGB}{160,32,240}
\definecolor{darkred}{RGB}{255,0,255}
\usepackage{stfloats}
\usepackage{bm}
\usepackage{multirow}
\usepackage{bbding} 
\usepackage{flushend}
\usepackage{hyperref}  

\begin{document}
	\title{Trajectory Design for UAV-Based Internet-of-Things Data Collection: A Deep Reinforcement Learning Approach}
	\author{Yang Wang, Zhen Gao, Jun Zhang, Xianbin Cao, Dezhi Zheng, Yue Gao, \IEEEmembership{Senior Member}, Derrick Wing Kwan Ng, \IEEEmembership{Fellow, IEEE}, and Marco Di Renzo, \IEEEmembership{Fellow, IEEE}
	\thanks{Y. Wang, Z. Gao, and J. Zhang are with the School of Information and Electronics, Beijing Institute of
		Technology, Beijing 100081, China (e-mail: 3120200843@bit.edu.cn; gaozhen16@bit.edu.cn; buaazhangjun@vip.sina.com).}
	\thanks{X. Cao is with the School of Electronic and Information Engineering, Beihang University, Beijing 100083, China (e-mail: xbcao@buaa.edu.cn).}
	\thanks{D. Zheng is with the Innovation Institute of Frontier Science and Technology, School of Instrumentation and Optoelectronic Engineering, Beihang University, Beijing 100083, China (e-mail: zhengdezhi@buaa.edu.cn).}
	\thanks{Y. Gao is with the Department of Electrical and Electronic Engineering, University of Surrey, Surrey GU2 7XH, U.K. (e-mail: yue.gao@ieee.org).}
	\thanks{D. W. K. Ng is with the School of Electrical Engineering and Telecommunications, University of New South Wales, Sydney, NSW 2025, Australia (e-mail: w.k.ng@unsw.edu.au).}
	\thanks{M. Di Renzo is with Universit{\'e} Paris-Saclay, CNRS, CentraleSup{\'e}lec, Laboratoire des Signaux et Syst{\`e}mes, 3 Rue Joliot-Curie, 91192 Gif-sur-Yvette, France (e-mail: marco.di-renzo@universite-paris-saclay.fr).}}
	
	\maketitle
	
	\begin{abstract}
		In this paper, we investigate an unmanned aerial vehicle (UAV)-assisted Internet-of-Things (IoT) system in a sophisticated three-dimensional (3D) environment, where the UAV's trajectory is optimized to efficiently collect data from multiple IoT ground nodes. Unlike existing approaches focusing only on a simplified two-dimensional scenario and the availability of perfect channel state information (CSI), this paper considers a practical 3D urban environment with imperfect CSI, where the UAV's trajectory is designed to minimize data collection completion time subject to practical throughput and flight movement constraints. Specifically, inspired from the state-of-the-art deep reinforcement learning approaches, we leverage the twin-delayed deep deterministic policy gradient (TD3) to design the UAV's trajectory and present a TD3-based trajectory design for completion time minimization (TD3-TDCTM) algorithm. In particular, we set an additional information, i.e., the merged pheromone, to represent the state information of UAV and environment as a reference of reward which facilitates the algorithm design. By taking the service statuses of IoT nodes, the UAV's position, and the merged pheromone as input, the proposed algorithm can continuously and adaptively learn how to adjust the UAV's movement strategy. By interacting with the external environment in the corresponding Markov decision process, the proposed algorithm can achieve a near-optimal navigation strategy. Our simulation results show the superiority of the proposed TD3-TDCTM algorithm over three conventional non-learning based baseline methods.
	\end{abstract}
	
	\section{Introduction}
	\label{sec:introduction}
	
	\IEEEPARstart{T}{}he unmanned aerial vehicle (UAV)-assisted communication paradigm is expected to play a pivotal role in the next-generation wireless communication systems for providing ubiquitous connectivity with a broader and deeper coverage\cite{summary,summary2}. Particularly, adopting UAVs as aerial mobile base stations (BSs) to collect data from distributed Internet-of-Things (IoT) ground nodes is anticipated to be a promising technology for realizing green communications\cite{IoT}. Compared with terrestrial BS-based IoT systems, the UAV-based aerial BS system has salient attributes, such as a high probability in establishing strong line-of-sight (LoS) channels to improve coverage, a flexible deployment and fast response for unexpected or limited-duration missions, and a dynamic three-dimensional (3D) placement and movement for improving spectral and energy efficiency, etc\cite{UAV_sum}.

	\subsection{Related Works}
	Due to the high maneuverability, UAVs can approach closely to the potential IoT nodes offering a low power consumption solution for the IoT nodes uploading data, where the UAVs' trajectory is essential to be carefully optimized. To date, there have been various related work investigating the trajectory design for UAV-assisted IoT data collection systems, where different optimization objectives, such as energy-efficiency\cite{efficient,efficient-2,efficient-3,efficient-4}, data collection rate\cite{Rate-1,Rate-2,Rate-3}, flight time\cite{time-1,time-2}, and age-of-information (AoI)\cite{AoI} have been considered. Specifically, the authors in \cite{efficient} considered to maximize the energy-efficiency of wireless sensor networks by optimizing the UAV's trajectory and sensors' wake-up schedule. Besides, in \cite{efficient-2}, to minimize the total transmit power of IoT nodes, the authors considered to optimize the UAVs' 3D placement, each IoT node's transmit power, and the association between IoT nodes and UAVs. Also, in \cite{efficient-3}, the authors studied the UAV's trajectory and the route design problem which the UAV adopted the hover-and-fly model for data collection. A suboptimal design was proposed to maximize the minimal residual energy of sensor nodes after data transmission. However, the above works mainly considered the sensors' energy consumption and failed to consider the UAV’s energy consumption, which is one of the important factors for determining the performance of UAV-assisted communication systems. In \cite{efficient-4}, the authors jointly optimized the UAV's trajectory, resource allocation strategy and the jamming policy for maximizing the UAV's energy efficiency.
	
	Generally, a UAV would cruise sufficiently close enough to each sensor for efficient data collection. However, this associated high energy consumption of the UAV would limit its maneuverability and endurance in long flight, which jeopardizes the total collection throughput of the whole network. Therefore, in additional to the energy-efficiency, there also have several works considering to optimize the data collection rate \cite{Rate-1,Rate-2,Rate-3,Rate-4}. For instance, the authors in \cite{Rate-1} jointly designed the sensors' transmission scheduling, power allocations, and the UAV's trajectory to maximize the minimum data collection rate from the ground sensors to a multi-antenna UAV. Also, in \cite{Rate-2}, the authors optimized the UAV's 3D trajectory to maximize the minimum average rate for data collection take into account the angle-dependent Rician fading channels. Besides, in \cite{Rate-3,Rate-4}, the authors considered to jointly design the UAV's trajectory and the wireless resource allocation optimization for maximizing the system throughput.
	
	On the other hand, there also have been several prior works studying the UAV-assisted data collection with flight time optimization\cite{time-1,time-2}. Specifically, the authors in \cite{time-1} and \cite{time-2} jointly designed the UAV's flight trajectory and wireless resource allocation/scheduling to minimize the mission completion time, where the sensors deployed in the one-dimensional (1D) and two-dimensional (2D) spaces are considered, respectively. Besides, there is a few work considering the AoI of the data collection. In \cite{AoI}, the authors considered to minimize the average AoI of the data collected from all sensors by jointly optimizing the UAV's trajectory, energy harvesting duration, and the time required for data collection at each sensor.
	\begin{table*}[htbp]
		\centering
		\caption{Comparison of related works with our work.}
		\resizebox{\textwidth}{!}{
			\begin{tabular}{|p{50pt}|p{130pt}|
					p{130pt}|p{130pt}|p{80pt}|p{40pt}|}
				\hline 
				References&Optimization objective&Optimization variables&Optimization method&Channel model&Number of UAVs\\
				\hline
				\cite{efficient}&Energy consumption of all sensors minimization&Sensors' wake-up schedule and UAV’s 2D trajectory&Successive convex optimization technique&Simplified LoS channel&Single\\
				\hline
				\cite{efficient-2}&Total transmit power of IoT devices minimization&3D placement of the UAVs, the association of devices, and uplink power control&Traditional convex optimization technique&Probabilistic LoS channel&Multiple\\
				\hline
				\cite{efficient-3}&Sensors' minimum residual energy maximization&UAV's hovering locations and 2D route&Voronoi diagram theory&Simplified LoS channel&Single\\
				\hline
				\cite{efficient-4}&UAV's energy-efficiency maximization&UAV's 2D trajectory, resource allocation, and jamming policy&Alternating optimization algorithm &Simplified LoS channel&Single\\
				\hline
				\cite{Rate-1}&Minimum data collection rate maximization&Sensors' transmission scheduling, power allocation, and UAV's 2D trajectory&Block coordinate descent and successive convex approximation&Simplified LoS channel&Single\\
				\hline
				\cite{Rate-2}&Minimum average data collection rate maximization&UAV's 3D trajectory&Block
				coordinate descent and successive convex approximation&Angle-dependent Rician fading channel&Single\\
				\hline
				\cite{Rate-3}&Uplink minimum throughput maximization&UAV's 2D trajectory and transmission resource allocation &Alternating optimization and successive
				convex programming&Simplified LoS channel&Single\\
				\hline
				\cite{Rate-4}&The sum throughput maximization&UAV's 3D trajectory and the transmit power and subcarrier allocation&Monotonic optimization and successive
				convex programming&Simplified LoS channel&Single\\
				\hline
				
				\cite{time-1}&UAV's total flight time minimization&UAV’s 1D speed and sensors' transmission power&Dynamic programming&Simplified LoS channel&Single\\
				\hline
				\cite{time-2}&Total mission time minimization&UAV’s 2D trajectory, fixed altitude, velocity, and scheduling&Block coordinate descent&Probabilistic LoS channel&Single\\
				\hline
				\cite{AoI}&Average age of information minimization&Time of energy transmission plus data collection, and UAV's 2D trajectory&Dynamic programming and ant colony heuristic algorithm\cite{ant}&Simplified LoS channel&Single\\
				\hline
				\cite{efficient_DRL1}&Energy-efficient maximization&UAV's 3D trajectory&Deep deterministic policy gradient algorithm (DDPG)\cite{DDPG}&Probabilistic LoS channel&Multiple\\
				\hline
				\cite{efficient_DRL2}&Energy-efficient maximization&UAV's 2D trajectory and charge station's 2D trajectory&Deep Q network (DQN) \cite{DQN}&None channel model&Single\\
				\hline
				\cite{efficient_DRL3}&Energy-efficient maximization&UAV's 2D trajectory&Multi-agent deep deterministic policy gradient (MADDPG) \cite{MADDPG}&None channel model&Multiple\\
				\hline
				\cite{CCU_RL}&Weighted sum of the mission completion time and the expected communication outage duration minimization&UAV's 2D trajectory&Temporal difference learning \cite{Bible}&Practical urban LoS channel&Single\\
				\hline
				\cite{AoI-RL}&Weighted sum of the age of information maximization&UAV’s 2D trajectory and transmission scheduling of the sensors&DQN&Simplified LoS channel&Single\\
				\hline
				Our work&Average mission completion time minimization&UAV's navigation altitude and UAV's 2D trajectory&Twin-delayed deep deterministic policy gradient algorithm (TD3) \cite{TD3}&Practical urban LoS channel&Single\\
				\hline
		\end{tabular}}
		\label{survey_tab}
	\end{table*}
	
	However, the above UAV trajectory designs based on conventional optimization solutions suffer from some critic limitations. First, the formulation of an optimization problem requires an accurate and tractable radio propagation model. For such reason, recent works, e.g., \cite{efficient,efficient-2,efficient-3,efficient-4,Rate-1,Rate-2,Rate-3,Rate-4,time-1,time-2,AoI}, have mostly adopted some statistical models such as the simplified LoS-dominated model, the probabilistic LoS model, and the angle-dependent Rician fading model. However, these models can only predict the performance in the average sense and cannot provide any performance guarantee for the local environment where the UAVs are actually deployed. Second, the offline optimization-based trajectory design assumed that perfect channel state information (CSI) can be explicitly derived from a specific radio propagation model. Unfortunately, perfect CSI in practice is often difficult to be acquired due to the uncertainty of the UAV's position and the time-varying dynamic of the communication environment. At last, most of these optimization problems in UAV-assist communication systems are highly non-convex and difficult to be solved efficiently. In addition, machine learning algorithms such as supervised learning and unsupervised learning require sufficient data samples in advance, which is unrealistic for decision-making problems. In contrast, deep reinforcement learning (DRL) has been serving as an efficient solution to tackle such a decision-making issue thanks to its essential traits, i.e., learning dynamically from the real world. Specifically, since there is no prior information about the environment model in the Markov decision process (MDP), the agent, i.e., the learning entity, would interact with the external environment, collect some samples in real-time, and then design the optimal strategy based on these samples space\cite{Bible}.
	
	Recently, there have been several research works leveraging DRL to optimize the UAV's trajectory for UAV-assisted IoT data collection systems. For example, the authors in \cite{efficient_DRL1} designed a DRL-based 3D continuous movement control algorithm to solve an trajectory optimization problem with the goal of maximizing the system energy efficiency. Also, in \cite{efficient_DRL2,efficient_DRL3}, the authors proposed a DRL-based UAV control method for maximizing the energy efficiency in mobile crowd sensing systems. Besides, in \cite{CCU_RL}, to minimize the weighted sum of the mission completion time and the expected communication outage duration, the authors focused on optimizing the UAV's trajectory based on DRL. Moreover, in \cite{AoI-RL}, the weighted sum of the AoI was adopted as the performance metric in UAV-assisted wireless powered IoT networks, where the UAV’s trajectory and transmission scheduling of the sensors were jointly optimized adopting a DRL-based approach. For clarity, the comparison of the related works\cite{efficient,efficient-2,efficient-3,efficient-4,Rate-1,Rate-2,Rate-3,Rate-4,time-1,time-2,AoI}, \cite{efficient_DRL1,efficient_DRL2,efficient_DRL3,CCU_RL,AoI-RL} is summarized in Table \ref{survey_tab}.
	
	\subsection{Motivations}
	Existing UAV trajectory design schemes, e.g., \cite{efficient,efficient-3,efficient-4,Rate-1,Rate-3,Rate-4,time-1,AoI,AoI-RL}, for IoT data collection systems rely on the existence of pure LoS channels. Although adopting the overly simplified LoS channel model facilitates the system performance analysis and resource allocation optimization, these aforementioned works fail to capture the dynamics in practical IoT communication environments. On the other hand, another well-known ground-air (G2A) channel model, known as probabilistic LoS model \cite{LAP}, has been adopted in\cite{efficient-2,time-2,efficient_DRL1} as an alternative. This channel model considers the existence of both LoS and non-line-of-sight (NLoS) links with certain probability, respectively, following a practically assumed distribution model. However, the existence of LoS link should depend on the actual environment, i.e., whether the direct connection between the UAV and IoT node is physically blocked by obstacles, rather than based on the probability model or simplified LoS channel model. To the best of our knowledge, only the authors in \cite{CCU_RL} considered a practical urban LoS model. Yet, its UAV's trajectory in \cite{CCU_RL} was designed for the single-sensor scenario, which do not applicable to the case of multiple-sensors and it only considered the optimization of the UAV in discrete horizontal direction. Therefore, there is few work considering such a practical urban LoS channel for UAV-assisted IoT data collection system, which motivates our research in this article.
	\subsection{Contributions}
	In this paper, we propose a trajectory design for the completion time minimization (TDCTM) in the UAV-assisted IoT data collection systems. Specifically, the UAV is employed to collect delay-tolerant data from multiple IoT ground nodes distributed in an urban scenario. In order to address the problems of practical urban LoS channel modeling and the availability of imperfect CSI problems, we leverage the twin-delayed deep deterministic policy gradient (TD3)\cite{TD3}, known as a state-of-the-art DRL approach, to design the UAV's trajectory for minimizing the mission completion time, subject to the constraints of throughput and flight movement. Our contributions are summarized as follows:
	\begin{itemize}
		\item{First, to minimize the mission completion time, we formulate the trajectory design for UAV-assisted IoT data collection systems. This problem, however, is generally intractable to be solved by conventional convex optimization methods due to a sophisticated distribution of buildings in relatively practical environment leading to imperfect CSI. To address this difficulty, we reformulate the original problem as an equivalent MDP, where DRL techniques can be applied to acquire a near-optimal solution for this problem.}
		\item{Next, to cope with the continuous control problem with an infinite action space, we propose a TD3-based TDCTM (TD3-TDCTM) algorithm realized by TD3, which can pilot the UAV agent continuously and adaptively learn how to adjust its movement strategy. Besides, inspired by ant colony algorithm\cite{ant}, we set an additional information, i.e., the merged pheromone, to represent the state information of UAV and environment which is adopted as a reward function. The proposed TD3-TDCTM algorithm requires only some commonly accessible information, i.e., the service statuses of the IoT nodes, the UAV's position, and the merged pheromone, where they can be obtained by minimum information exchange between the agent and the environment.}
		\item{Then, to facilitate the convergence of TD3-TDCTM: i) we propose an {\it information-enhancement} technique that a self-defined pheromone of the UAV is added to the state for enhancing the learning efficiency; ii) considering the sparse reward of the original problem, we propose a {\it reward-shaping} mechanism, which can transform the original sparse rewards as dense rewards; iii) to address the convergence of Q-loss, we integrate the techniques including {\it dimension-spread} and {\it done-or-terminated}. Also, we conduct extensive simulations to verify the effectiveness of the above four techniques.}
		\item{Last, we compare the proposed algorithm with various baseline methods.
		Results show that the proposed algorithm consistently outperforms the others. Also, we find an appropriate set of hyperparameters that achieve good performance in terms of discount factor, neuron number, and experience replay buffer size by extensive simulations.}
	\end{itemize}
	
	The remainder of this article is organized as follows. In Section \uppercase\expandafter{\romannumeral2}, the UAV-assisted IoT communication system model is presented. In Section \uppercase\expandafter{\romannumeral3}-A, the UAV data collection problem is formulated as a non-convex optimization problem. Section \uppercase\expandafter{\romannumeral3}-B provides a brief introduction of TD3. As for the rest parts of Section \uppercase\expandafter{\romannumeral3}, the TD3-TDCTM algorithm is described in details. Then, simulation results are presented in Section \uppercase\expandafter{\romannumeral4} to evaluate the performance of the proposed algorithm, as well as its robustness and convergence against various parameters. Finally, we conclude the paper in Section \uppercase\expandafter{\romannumeral5}.
	
	{\it Notations:} In this paper, scalars are denoted by italic letters, and vectors are denoted by boldface letters. The Euclidean norm of a vector is denoted by $\left \| \cdot\right\|$, $\left\{\cdot \right\}$ denotes an array, $\mathbb{E}\left[\cdot\right]$ denotes the statistical expectation, $\nabla$ denotes the gradient of function, and ${\rm clip}\left(\cdot \right)$ denotes the boundary clipping function. The distribution of a Gaussian random noise with mean $\mu$ and covariance $\sigma^2$ is denoted by $\mathcal{N}\left(\mu,\sigma^2\right)$.
	
	\section{System Model}\label{II}
\begin{figure}[htbp]
		\centering{\includegraphics[width=\columnwidth,keepaspectratio]{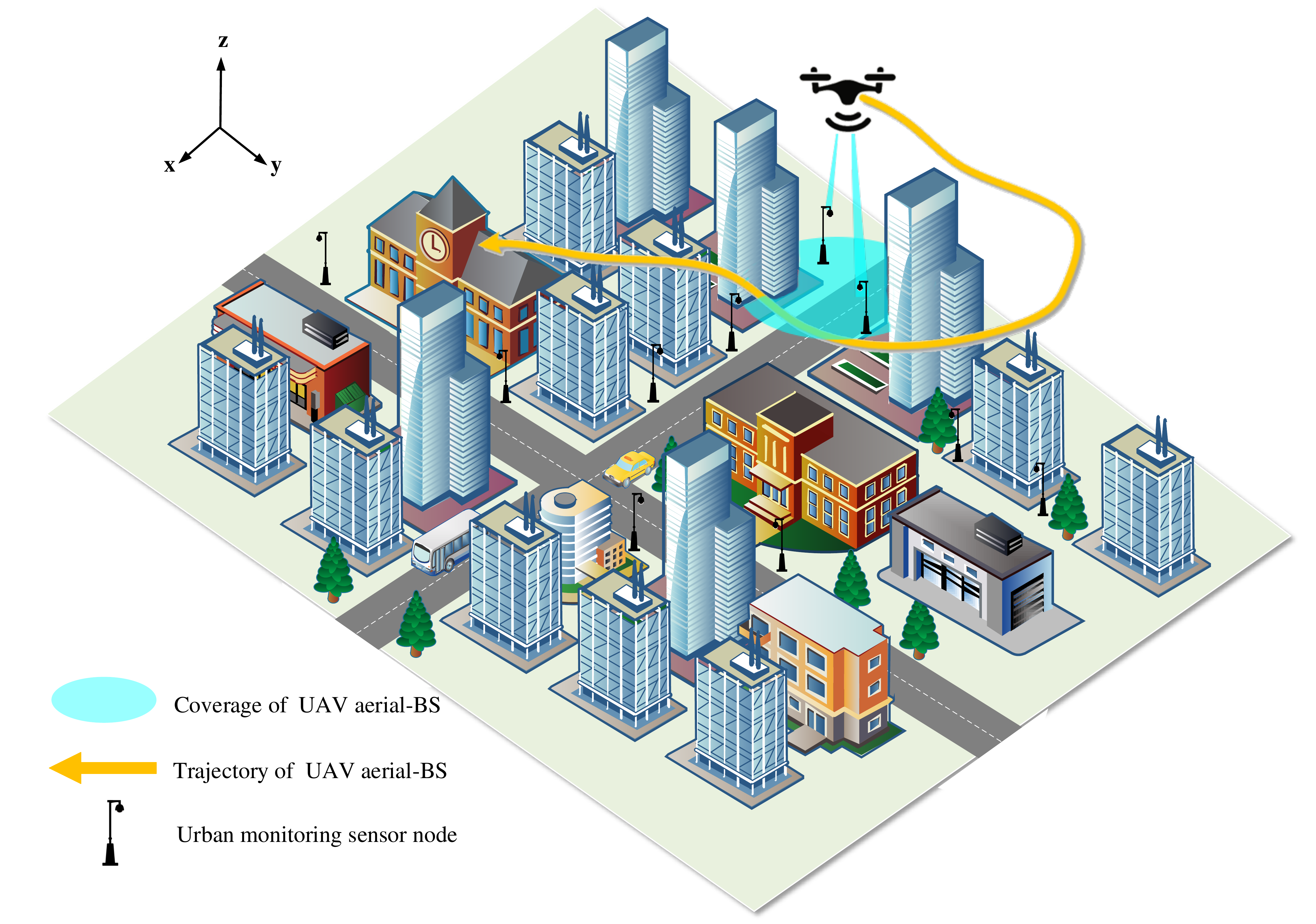}}
		\captionsetup{font={footnotesize}, name={Fig.},labelsep=period}
		\caption{UAV-assisted IoT data collection system.}
		\label{fig:scenario}
	\end{figure}

	As shown in Fig. \ref{fig:scenario}, we consider a UAV-assisted IoT data collection system, where the UAV is dispatched to collect delay-tolerant data from a large number of distributed IoT ground nodes, e.g., the wireless sensors in the applications of smart city monitoring, traffic flow, health monitoring, and etc. We assume that both the UAV and the IoT ground nodes employ single omni-directional antenna and $K$ IoT nodes are randomly distributed in a given geographical region of $D\times D$ m$^2$. The positions of the $k$-th IoT node and the UAV are denoted by $\boldsymbol{w}_k=[\bar{x}_k,\bar{y}_k,0]\in\mathbb{R}^{3}$ and $\boldsymbol{q}(t)=[x_{t},y_{t},H]\in\mathbb{R}^{3}$, $0\le t\le T$, respectively, where $(\bar{x}_k,\bar{y}_k)$ denotes the horizontal coordinate of the $k$-th IoT node, $(x_{t},y_{t})$ is the UAV location projected on the horizontal plane, $H$ is the UAV altitude, and $T$ is the mission execution duration. The $K$ IoT nodes' locations are assumed to be fixed for designing the UAV's trajectory and performing data collection. 
	
	As mentioned in Table \ref{survey_tab}, most existing works adopt the simplified LoS channel or the statistic channel model (i.e., the probabilistic LoS channel) to model the G2A link. By contrast, we consider a more practical G2A channel model, which can be characterized by large-scale fading and small-scale fading which are calculated based on a simulated 3D map by taking into account the existence of buildings as propagation scatterers. Specifically, the locations and the heights of the buildings are generated according to a statistical model recommended by the international telecommunication union (ITU)\cite{ITU}. In this model, there are three parameters to characterize an urban environment: $\alpha$ is the ratio of a land area covered by buildings to the total land area; $\beta$ is the average number of buildings per square kilometer; and $\lambda$ is the mean value of a Rayleigh distribution which represents the building height distribution.

	Given a specific area with the simulated building location and height, we can accurately determine whether there is a LoS link between the UAV and the $k$-th IoT node by checking whether the direct communication link between them is blocked by a building. Thus, the large-scale fading of the G2A channel associated with the $k$-th IoT node can be expressed as \cite{LAP}
	\begin{equation} {\rm PL}_{k}(t)=\begin{cases}L_{k}^{\rm{FS}}(t)+\eta_{\rm LoS},\\
	L_{k}^{\rm{FS}}(t)+\eta_{ \rm NLoS},
	\end{cases}
	\label{Eq1}
	\end{equation}
	where ${L_{k}^{\rm{FS}}}(t)=20\log_{10} d_{k}(t)+20\log_{10} f_c+20\log_{10} \left(\frac{4\pi}{c}\right)$ represents the free space pathloss between the UAV and the $k$-th IoT node, $d_k(t)=\left \| \boldsymbol{q}(t)-\boldsymbol{w}_k \right \|$ denotes the distance from the UAV to the $k$-th IoT node, $f_c$ denotes the carrier frequency, and $c$ represents the velocity of light. Besides, $\eta_{\rm LoS} $ and $\eta_{\rm NLoS}$ represent the propagation loss of the LoS and NLoS links, respectively\footnote{The above pathloss expressions are all in dB.}.
	
	The small-scale fading coefficient $\tilde{h}_{k}(t)$ is assumed to be Rayleigh fading for the NLoS case and Rician fading with 15 dB Rician factor for the LoS case, respectively. Furthermore, the Doppler effect caused by the UAV mobility is assumed to be well estimated and then compensated at the receiver by using existing advanced compensation algorithms\cite{Doppler}. Thus, the channel gain from the UAV to the $k$-th IoT node can be expressed as
	\begin{equation}
	h_{k}(t)= 10^{-{\rm PL}_{k}\left(t\right)/20} \tilde{h}_{k}\left(t\right).\label{Eq2}
	\end{equation}
	
	\section{Proposed DRL-Based TD3-TDCTM Scheme}\label{III}
	In this section, we formulate the UAV's trajectory optimization problem and reformulated it as a MDP structure, which is the common representation of DRL framework. Furthermore, we propose the TD3-TDCTM algorithm with three tricks of facilitating convergence for minimizing the mission completion time.
	\subsection{Problem Formulation}\label{A}
	To make the UAV's trajectory optimization problem tractable, the continuous time domain is discretized into $N$ time steps with unequal-length duration $\delta_n$, $n\in\{0,1,\ldots,N\}$, and a data collection task is performed within a series of time steps, i.e., $\{\delta_0,\delta_1,\ldots,\delta_N\}$\cite{efficient}. In addition, we consider that each time step consists of two parts, i.e., $\delta_n=\delta_{\rm ft}+\delta_{{\rm ht},n}$, where $\delta_{\rm ft}$ is the fixed flight time and $\delta_{{\rm ht},n}$ is the hovering time for data collection. If there is no active IoT node in the current position, the UAV would skip hovering and directly executes the next time step, i.e., $\delta_{{\rm ht},n}=0$ s. During each time step, the UAV mobility strategy can be expressed as
	\begin{equation}
	x_{n+1}= x_{n}+m_n\cos(\bar{\theta}_{n}),\label{Eq3}
	\end{equation}
	\begin{equation}
	y_{n+1}= y_{n}+m_n\sin(\bar{\theta}_{n}),\label{Eq4}
	\end{equation}
	where $m_n=\delta_{\rm ft}{\upsilon_n}$ represents the moving distance of the UAV in the $n$-th time step, $\upsilon_n\in[0,\upsilon_{\rm max}]$ denotes the average flight speed in the $n$-th time step, $\upsilon_{\rm max}$ denotes the maximum cruising speed in each time step, and $\bar{\theta}_{n}\in (0,2\pi] $ denotes the horizontal direction of the UAV in the $xy$-plane with respect to the $x$-axis. 
	
	
	We assume that only when IoT nodes are waken up by the UAV, they can begin to upload data with a constant transmission power $P_{\rm Tx}$, otherwise they continues to stay in the silent mode for energy saving. In the $n$-th time step, the corresponding uplink signal-to-noise ratio (SNR) between the $k$-th IoT node and the UAV can be expressed as
	\begin{equation}
	\rho_{k,n}=\frac{P_{\rm Tx}{\left\vert h_{k,n} \right\vert}^2}{P_N},\label{Eq5}
	\end{equation}
	where $|h_{k,n}|^2$ is the channel power gain at the hover stage of the $n$-th time step and $P_N$ represents the power of the additive white Gaussian noise (AWGN) at the UAV receiver. For the uplink data collection service associated with the $k$-th IoT node, we set a pre-defined SNR threshold $\rho_{\rm th}$, where the $k$-th IoT node can be waken up and served by the UAV if and only if $\rho_{k,n} \ge \rho_{\rm th}$. Furthermore, we define the indicator function as
	\begin{equation}
	b_{k,n}=\begin{cases}1,\quad {\rm if}\, \rho_{k,n}\ge \rho_{\rm th}, \\
	0,\quad {\rm otherwise},
	\end{cases}
	\label{eq-indict}
	\end{equation}
	to indicate whether the $k$-th IoT node can be satisfied with the SNR requirement by the UAV in the $n$-th time step. Due to the assumption that each IoT node only can be served at most once in one realization, we define the following indicator function of the $k$-th IoT node as 
	\begin{equation}
	\tilde{b}_{k,n}=\begin{cases}1,\quad {\rm if}\, b_{k,n}=1, {\rm and}\, c_{k,n}=0, \\
	0,\quad {\rm otherwise},
	\end{cases}
	\label{eq-indict2}
	\end{equation}
	where $c_{k,n}\in\left\{0,1\right\}$ is a binary variable to indicate whether the $k$-th IoT node has been served by the UAV. For simplicity, we assume that an orthogonal frequency division multiple access (OFDMA) is utilized to allow the simultaneous data collection from at most $K_{\rm up}$ IoT nodes, i.e., each IoT node satisfying upload requirements would be allocated with bandwidth $W$, thus the inter-user-interference can be neglected. Therefore, the constraint of $\tilde{b}_{k,n}$ can be expressed as
	\begin{equation}
	\sum_{k=1}^{K}\tilde{b}_{k,n}\le K_{\rm up},
	\end{equation}
	where if the number of wake-up IoT nodes exceeds the maximum access number $K_{\rm up}$, the system will select the first $K_{\rm up}$ IoT nodes with the large SNR as the serving objects, and set other IoT nodes to be silent, i.e., $\tilde{b}_{k,n}=0$.
	Besides, we define the serving flag $c_{k,n}$ as
	\begin{equation}
	c_{k,n/0}=\min\left\{\sum_{i=0}^n \tilde{b}_{k,i}, 1\right\},\,c_{k,0}=0,\label{eq-indict3}
	\end{equation}
	where if $c_{k,n}=1$, the $k$-th IoT node has been served during the mission; otherwise, the $k$-th IoT node has not been served. Then, the transmission rate between the UAV and the $k$-th IoT node can be expressed as
	\begin{equation}
	R_{k,n}=\tilde{b}_{k,n}W\log_2\left(1+\rho_{k,n}\right),
	\end{equation}
	where only when the $k$-th IoT node is served in the $n$-th time step, the data of the $k$-th IoT node can be uploaded. Thus, the hovering time of UAV, which equals to the maximum upload data duration from the served IoT nodes in the $n$-th time step, can be expressed as
	\begin{equation}
	\delta_{{\rm ht},n}= \max_{k\in\{1,\ldots,K\}}\left\{ \frac{\tilde{b}_{k,n}D_{\rm file}}{R_{k,n}+\kappa}\right\},\label{eq-ht}
	\end{equation}
	where $D_{\rm file}$ denotes the information file size of the $k$-th IoT node and $\kappa$ is the value preventing the denominator from being zero.
	The completion criterion of the data collection mission is that the data of all IoT nodes has been collected, which can be expressed as
	\begin{equation}
	\sum_{k=1}^K c_{k,N}= K.\label{eq-comp}
	\end{equation}
	
	Thus, the problem to minimize the mission completion time via trajectory optimization can be formulated as
	\begin{equation}
	\begin{array}{cll}
	\displaystyle\mathop{\mathrm{minimize}}\limits_{\left\{\upsilon_n, \bar{\theta}_{n}\right\},\left\{x_n,y_n\right\},N}\quad & \sum_{n=0}^{N}\delta_n \\ 
	{\rm s.t.} \quad & \sum_{k=1}^{K}\tilde{b}_{k,n}\le K_{\rm up},\forall n,\\
	& c_{k,n}=\min\left\{\sum_{i=0}^n \tilde{b}_{k,i}, 1\right\}, \forall n,k,\\
	& \sum_{k=1}^K c_{k,N}= K,\\
	& 0 \le {\upsilon_n} \le \upsilon_{\rm max}, \forall n,\\
	& 0 \le \bar{\theta}_n \le 2\pi, \forall n,\\
	& 0 \le x_n \le D, \forall n, \\
	& 0 \le y_n \le D, \forall n. \\
	\end{array}
	\label{Eq8}
	\end{equation}
	
	It is noteworthy that the above optimization problem is a mixed-integer non-convex problem, which is known to be NP-hard. Moreover, in the considered scenario, both the large-scale fading and small-scale fading depend on the instantaneous locations of both the UAV and the IoT nodes as well as the surrounding buildings, so that it is intractable to solve the above problem by employing traditional optimization methods, such as constraint programming and mixed integer programming \cite{Convex}. In contrast, DRL has been demonstrated to be an efficient approach for handling sophisticated control problems in high-dimensional continuous spaces\cite{TD3}. Hence, in the following subsection, we propose a DRL-based solution to tackle this challenging control problem.
	\subsection{Preliminaries}\label{TD3}
	Reinforcement learning (RL) considers the paradigm of an agent interacting with its environment with the aim of learning reward-maximizing policy\cite{Bible}. Specifically, RL can be used to address a MDP problem with 4-tuple $\left \langle \mathcal S, \mathcal A, \mathcal P, \mathcal R \right \rangle$, where $\mathcal S$ is the state space, $\mathcal A$ is the action space, $\mathcal P$ is the state transition probability, and $\mathcal R$ is the reward function. At each discrete time step $n$, with a given state $s\in \mathcal{S}$, the agent selects action $a\in \mathcal{A}$ with respect to its policy $\pi$, and receives a reward $r$. The return is defined as $R_n=\sum_{i=n}^{N}\gamma^{i-n}r\left(s_i,a_i\right)$, where $\gamma$ is a discount factor determining the priority of short-term rewards.
	
	DRL can be considered as the ``deep" version of RL, which uses multiple DNNs as the approximator of the Q-value function $Q\left(s,a\right)=\mathbb{E}\left[R_n|s,a\right]$. Here, $Q\left(s,a\right)$ is the expected return when performing action $a$ in state $s$. In deep deterministic policy gradient (DDPG) algorithm\cite{DDPG}, the Q-value approximator $Q_{\theta}\left(s,a\right)$ with parameters $\theta$ can be updated by minimizing the following loss function
	\begin{equation}
	L\left(\theta\right)=\mathbb{E}\left[\left(y-Q_{\theta}\left(s,a\right)\right)^2\right],
	\end{equation}
	where $y$ is the target value, which can be estimated by
	\begin{equation}
	y=r+\gamma Q_{\theta^\prime}\left(s^{\prime},a^{\prime}\right),a^{\prime}\sim \pi_{\phi^{\prime}}\left(s^{\prime}\right),
	\end{equation}
	where $s^{\prime}$ is the next state, $a^{\prime}$ is an action selected from a target actor network $\pi_{\phi^{\prime}}$, and $Q_{\theta^\prime}$ is a target network to maintain a fixed objective $y$ over multiple updates. The policy can be updated through the deterministic policy gradient algorithm which is given by
	\begin{equation}
	{\nabla }_{\phi } {J(\phi)}=\mathbb {E}\left[{\nabla }_{ {a}}  {Q}_{ {\theta}}(s,a) {\vert }_{ {a=}\pi _{\phi } {(s)}}{ {\nabla }}_{\phi }\pi _{ {\phi }}(s)\right].
	\end{equation}
	
	As a realization of the celebrated actor-critic algorithm, DDPG can achieve good performance in many applications. Yet, it has a fatal shortcoming, that is, the problem of overestimation, which will lead to a cumulative error. In particular, this kind of error would cause the originally poor state to admit a high Q value leading to a highly non-optimal strategy. In order to address this issue, TD3 was proposed by\cite{TD3}, which considers the interplay between policy and value updates in the approximation errors of function. Based on DDPG, TD3 applies the following three techniques to address the overestimation bias problem and achieves better performance over its counterparts, e.g., proximal policy optimization \cite{PPO} and soft actor-critic \cite{SAC}.
	\begin{figure}[htbp]
		\centering{\includegraphics[width=\columnwidth,keepaspectratio]{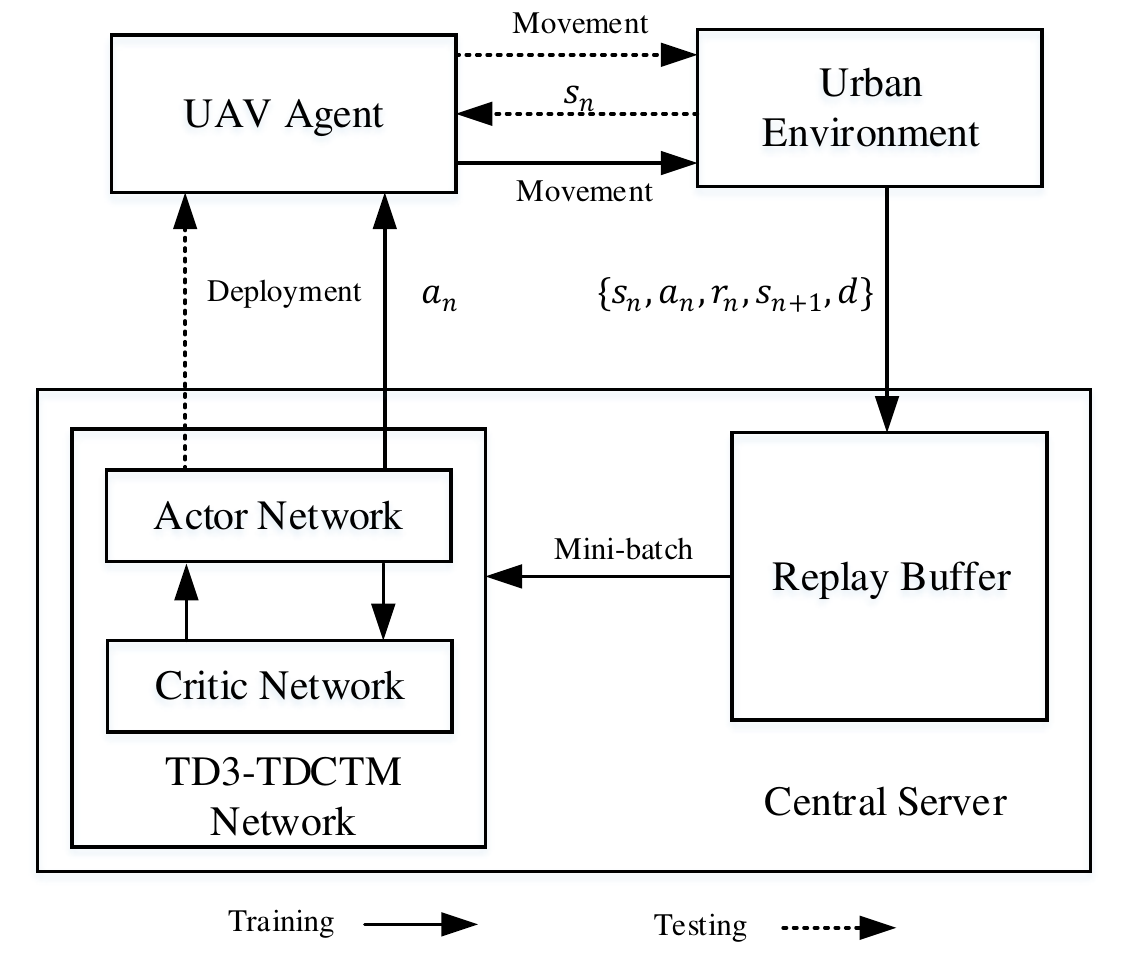}}
		\captionsetup{font={footnotesize}, name={Fig.},labelsep=period}
		\caption{TD3-based trajectory design for UAV-assisted IoT data collection system.}
		\label{fig:net_uav}
		\vspace{-6mm}
	\end{figure}

	The first technique is {\it clipped double Q-learning}. TD3 learns two Q-functions instead of one (hence ``twin"), i.e., $Q_{\theta_1}$ and $Q_{\theta_2}$. Therefore, critic network and critic target network both contain two independent Q networks, respectively. In each update, the Q target network with smaller Q-value is chosen as the Q target. Specifically, the target update of the clipped double Q-learning can be expressed as
	\begin{equation}
	y = r+\gamma \min_{i=1,2} \left\{Q_{\theta^\prime_i}\left(s^{\prime},\tilde{a}\right)\right\},
	\end{equation}
	where $r$ is a reward that we designate and $\tilde{a}$ is the output of target policy. 
	
	The second technique is {\it delayed policy updates}. TD3 updates the actor network and the target network less frequently than the critic networks. That is, the model does not update the policy unless the model's value function has changed sufficiently. These less frequent policy updates will lead to a lower variance in the value estimation and thus should result in a better policy. This allows the critic network to become more stable and reduce errors before it is used to update the actor network.

	The third technique is {\it target policy smoothing regularization}. A concern with deterministic policies is that they may overfit to narrow peaks in the value estimate. When updating the critic, a learning target using a deterministic policy is highly susceptible to inaccuracies induced by approximation errors of function, which can increase the variance of the target. This induced variance can be reduced through regularization. TD3 adds noise (i.e., clipped random noise) to the target action and averages over mini-batches with size $B$ for smoothing the value estimate, as shown below, 
	\begin{equation}
	\tilde{a}=\pi_{\phi^{\prime}}\left(s^{\prime}\right)+\epsilon, \epsilon\sim \rm{clip}\left ({\mathcal{N}\left ({\mathrm {0,\tilde{\omega}} }\right)\mathrm {,-c,c} }\right),
	\end{equation}
	where the added noise is clipped to keep the target close to the original action. The TD3 algorithm's update principles are similar to the DDPG algorithm, which can be expressed as
	\begin{align}
	\theta_{{i}}&\leftarrow {\mathrm{arg\, min}}_{\theta_{i}} {B}^{\mathrm { -1}}\sum (y- {Q}_{ {\theta }_{ {i}}}(s,a))^{2}, \label{L}\\
	{\nabla }_{\phi } {J(\phi)}&= {B}^{-1}\sum {\nabla }_{ {a}}  {Q}_{ {\theta_ 1}}(s,a) {\vert }_{ {a=}\pi _{\phi } {(s)}}{ {\nabla }}_{\phi }\pi _{{\phi }}(s). \label{J}
	\end{align}
	\subsection{MDP Formulation}\label{mdp}
	Combing the optimization problem formulated in Section \uppercase\expandafter{\romannumeral3}-A and the state-of-the-art DRL approach, TD3 in Section \uppercase\expandafter{\romannumeral3}-B, we reformulate the original problem as a MDP structure so that DRL algorithm can be applied. As shown in Fig. \ref{fig:net_uav}, the UAV is treated as an agent. During the training process, the TD3-TDCTM network and replay buffer are deployed on a central server. The replay buffer regularly collects the current state information of the interaction between the UAV and the environment, then the TD3-TDCTM network selects a better strategy to control the flight path of the UAV agent based on the historical states and rewards. When the actor network has been trained well, we will deploy it on the UAV agent during the testing stage. Then the UAV receives the state information of the environment and obtains the flight action command directly through the actor network. This process repeats until the mission is completed. Therefore, we first define the following state, action, and reward for UAV trajectory design problem in the following.

	1) State $s_n$ in the $n$-th time step, $\forall\, n$:
	\begin{itemize}
		\item{$b_{k, n}\in\{0, 1\}$: the coverage indicator of the $k$-th IoT node in the $n$-th time step. If $b_{k, n}=1$, the $k$-th IoT node satisfies the SNR requirement; otherwise, the $k$-th IoT node does not satisfy the requirement. Hence, our proposed DRL-based trajectory design only needs the imperfect CSI, i.e., SNR information.}
		\item{$c_{k,n}\in\{0, 1\}$: the serving flag of the $k$-th IoT node during the whole mission. If $c_{k,n}=1$, the $k$-th IoT node has been served during the mission; otherwise, the $k$-th IoT node has not been served.}
		\item{$[x_n,y_n]$: the two-dimensional coordinate of the UAV in a given region.}
		\item{$\zeta_n$: the pheromone of the UAV, which can be expressed as 
			\begin{equation}
			\zeta_n = \zeta_{n-1} + K_{{\rm cov},n}\cdot \kappa_{\rm cov}-\kappa_{\rm dis}-P_{\rm ob},
			\end{equation}
			where $\zeta_{n-1}$ is the remaining pheromone in the $\left(n-1\right)$-th time step, $K_{{\rm cov},n} = \sum_{k=1}^{K}\tilde{b}_{k,n}$ is the number of IoT nodes served by the UAV in the $n$-th time step, $\kappa_{\rm cov}$ is a positive constant that is used to express the captured pheromone per IoT node, $\kappa_{\rm dis}$ is a positive constant expressing the lost pheromone, and $P_{\rm ob}$ is a penalty when an action causes the boundary violation of the UAV.}
	\end{itemize}
	Formally, $s_n=[b_{1,n},\cdots, b_{K,n}; c_{1,n},\cdots, c_{K,n}; x_n, y_n; \zeta_n]$  is the complete representation of the $n$-th state, which has a cardinality equal to $2K+3$. In state $s_n$, both $b_{k,n}$ and $c_{k,n}$ reflect the data collection situation of the $k$-th IoT node; $[x_n,y_n]$ represents the UAV's movement status; and $\zeta_n$ denotes the merged information between environment and UAV agent during the mission, which can be regarded as an additional information to enhance the decision efficiency.
	
	\begin{algorithm}[tp!] 
		\caption{TD3-TDCTM}
		\label{alg 1}
		\begin{algorithmic}[1] 
			\STATE Initialize critic networks $Q_{\theta_1}$, $Q_{\theta_2}$, and actor network $\pi_{\phi}$ with random parameters $\theta_1$, $\theta_2$, and $\phi$
			\STATE Initialize target networks $\theta_1^{\prime} \leftarrow \theta_1$, $\theta_2^{\prime} \leftarrow \theta_2$, and $\phi^{\prime} \leftarrow \phi$ 
			\STATE Initialize experience replay buffer $R$
			\FOR{episode $=0$ to $M$} 
			\STATE Initialize the environment, receive an initial state $s_0$, let the time step $n=0$ and the terminated flag $d=0$
			\REPEAT 
			\STATE Select an action $a_n={\pi_{\phi} \left(s_n \right)}+\sigma\epsilon$, where $\epsilon$ is a Gaussian noise and $\sigma$ is a decay constant, and observe a reward $r_n=r_{\rm tanh}\left(\zeta_n\right)$ and a new state $s_{n+1}$
			\IF {the UAV flies over the border}
			\STATE $\zeta_n=\zeta_n-P_{\rm ob}$, where $P_{\rm ob}$ is a given penalty. Meanwhile, the movement of the UAV is canceled and update $r_n$, $s_{n+1}$ accordingly
			\ENDIF
			\IF{the UAV completes the data collection task, i.e., $\sum_{k=1}^K c_{k,n}=K$} 
			\STATE $r_n=r_n+N_{\rm re}$, and the episode is terminated in advance, i.e., $d=1$
			\ENDIF
			\STATE Store the transition $\left(s_n, a_n, r_n, s_{n+1},d\right)$ in $R$
			\IF{$R>$ 2,000}
			\STATE Sample mini-batch of $B$ transitions $\left(s, a, r, s^{\prime},d\right)$ from $R$
			\STATE $\tilde{a} \leftarrow \pi_{\phi^{\prime}} \left(s^{\prime}\right)+\epsilon$, $\epsilon \sim {\rm clip}\left(\mathcal{N}\left(0,\tilde{\omega}\right), -c, c\right)$
			\STATE Set $y \leftarrow r+\left(1-d\right)\cdot\gamma \min_{i=1,2} \left\{Q_{\theta^\prime_i}\left(s^{\prime},\tilde{a}\right)\right\}$
			\STATE Update critics by minimizing the loss:
			\STATE $\theta_i \leftarrow {\rm arg\,min}_{\theta_i} B^{-1}\begin{matrix} \sum \left(y-Q_{\theta_i}\left(s, a\right)\right)^2 \end{matrix}$
			\STATE Update the actor policy $\phi$ by the deterministic policy gradient:
			\STATE $\nabla{\phi}J\left(\phi\right)= B^{-1}\begin{matrix} \sum \nabla_a Q_{\theta_1}\left(s, a\right)|_{a=\pi_{\phi}\left(s\right)}\nabla_{\phi}\pi_{\phi}\left(s\right)\end{matrix}$
			\STATE Update target networks:
			\STATE $\theta_i^{\prime} \leftarrow \tau \theta_i+\left(1-\tau\right)\theta_i^{\prime}$ 
			\STATE  $\phi^{\prime} \leftarrow \tau \phi+\left(1-\tau\right)\phi^{\prime}$ 
			\ENDIF
			\STATE Update $n\leftarrow n+1$
			\UNTIL{$n=N_{\max}$ or $\sum_{k=1}^K c_{k,n}=K$}
			\ENDFOR
		\end{algorithmic} 
	\end{algorithm}
	
	2) Action $a_n$ in the $n$-th time step, $\forall\, n$: 
	\begin{itemize}
		\item $\bar{\theta}_n\in(0,2\pi]$: the horizontal direction of the UAV in the next time step.
		\item $\upsilon_n\in[0,\upsilon_{\rm max}]$: the flying speed of the UAV in the next time step.
	\end{itemize}
	Formally, the action is defined as $a_n=[\bar{\theta}_n, \upsilon_n]$. Since both action variables take continuous values, the UAV's trajectory optimization is a continuous control problem.
	
	3) Reward $r_n$ in the $n$-th time step, $\forall\, n$.
	For the above data collection mission, the UAV agent can not obtain a positive reward until it completes the data collection from all IoT nodes within the specified time step, i.e., there is no reward for every step in the intermediate process. Furthermore, at the beginning of training, the agent's strategy is random and the reward acquisition needs a series of complex operations. Therefore, the data collection mission is a sparse rewards problem\cite{Bible}. When we directly exploit reinforcement learning algorithm to optimize such a problem, the training difficulty of the original algorithm will increase exponentially as the number of IoT nodes increases, and the convergence cannot be guaranteed. To overcome this issue, we propose a {\it reward-shaping} mechanism, which can transform the original sparse rewards into dense rewards. Specifically, the reward design is defined as
	\begin{equation} r_{n}=
	\begin{cases}
	r_{\rm tanh}\left(\zeta_n\right)+N_{\rm re}, &{\rm if}\, \sum_{k=1}^K c_{k,n}=K, \\
	r_{\rm tanh}\left(\zeta_n\right), &{\rm otherwise},
	\end{cases}
	\label{Eq9}
	\end{equation}
	where $r_{\rm tanh}\left(\zeta_n\right)=\frac{2}{1+{\rm exp}\left(-\zeta_n/\left(K\cdot \kappa_{\rm cov}\right)\right)}-1$ is a shaped reward function of the pheromone $\zeta_n$. Besides, $r_{\rm tanh}\left(\cdot\right)$ approximates ${\rm tanh}\left(\cdot\right)$ function, but the gradient is smoother than the latter. Due to the change of pheromone $\zeta_n$, the UAV agent can obtain dense rewards within the exploration stage. Furthermore, the gradient information of reward function can accelerate the convergence of the algorithm. Besides, the UAV would obtain a remaining time reward $N_{\rm re}=N_{\max}-n$ at the mission completion time step, which thus encourages the UAV to complete the data collection mission as soon as possible.
	
	\subsection{TD3-Based UAV Trajectory Design}\label{proposed}
	To solve the UAV's trajectory design issue, we propose a TD3-TDCTM algorithm, whose pseudocode is listed in Algorithm 1. As mentioned above, since the UAV's trajectory design is a continuous control task, we adopt TD3, a state-of-the-art actor-critic algorithm, as the starting point of our design, whose basic idea has been introduced in Section \uppercase\expandafter{\romannumeral3}-B. In the following, we utilize {\it information-enhancement}, {\it dimension-spread}, and {\it done-or-terminated} tricks to stabilize the training process of TD3, whose details are illustrated below:

	1) {\it Information-enhancement}: Inspired by ant colony algorithm\cite{ant}, we set an additional information, i.e., the merged pheromone $\zeta_n$, as part of the state to enhance the learning efficiency. We assume that each IoT node contains some pheromones, which can also be represented as some special data to be collected. During the cruising of the UAV, the data of active IoT nodes will be collected and the pheromones on the IoT nodes also will be transferred to the UAV. At the same time, pheromones on the UAV will evaporate continuously and more pheromones will evaporate when the UAV's movement violates the boundary. The pheromone represents the fusion information of UAV and environment, which also serves as a reference for reward design. Besides, the pheromone concentration determines the reward, so as to pilot the UAV agent to explore environment better.
	
	\begin{figure*}
		\centering
		\includegraphics[width=\textwidth, keepaspectratio]
		{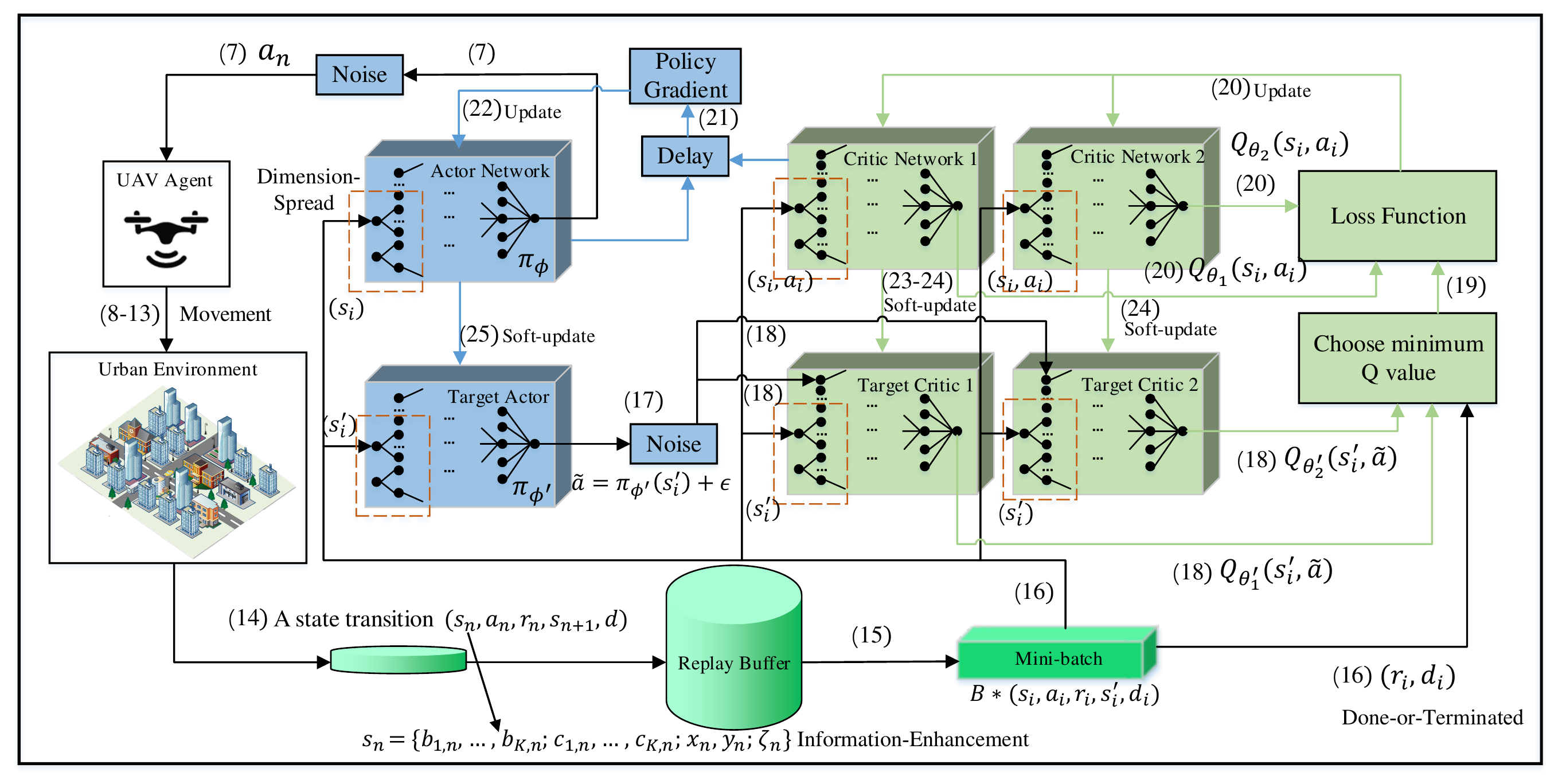}
		\captionsetup{font={footnotesize}, name={Fig.},labelsep=period}
		\caption{The TD3-TDCTM network framework proposed for the UAV-assisted IoT data collection system, where each step of Algorithm 1 is marked as a decimal number.}
		\label{net_framework}
	\end{figure*}
	
	2) {\it Dimension-spread}: According to the definition of state in the MDP formulation, most of the state dimensions are related to the coverage indicator of IoT nodes, while only two dimensions are related to location of the UAV and only one dimension is related to the pheromone of the UAV. Obviously, there is a dimension imbalance problem and we establish a pre-spread network to spread these state dimensions such that they are comparable to the coverage indicator dimensions \cite{Spread}. To this end, the low-dimensional states including the UAV's location and pheromone are first connected to a dense network with $2K$ neurons for extending the dimension to $2K$. Then these spread states and coverage indicator states are concatenated as the input of the actor and critic networks.
	
	3) {\it Done-or-terminated}: There are two situations to trigger ``done" in the environment: reaching the maximum time steps and completing the task. According to Bellman equation, the target value used for critic update is calculated as follow
	\begin{equation}
	Q\left(s,a\right)\approx r+\left(1-I_{\rm done}\right)\cdot \gamma \min_{i=1,2}\left\{ Q_{\theta^\prime_i}\left(s^{\prime},\tilde{a}\right)\right\},
	\end{equation}
	where $I_{\rm done}\in\left\{0,1\right\}$ is a binary variable to indicate whether the episode is over or not, i.e., ``done". However, it is incorrect to set ``done" when reaching the maximum time steps of the episode, because this episode is artificially terminated and the future Q value is abandoned. In fact, if the environment continues to run, the future Q value is not zero. Note that only when the UAV completes the task and terminates, the future Q value can be set to zero\footnote{This is because the target of the framework is to train the UAV to complete the data collection task. Once the task is completed, the episode will truly end and there will be no reward signal.}. If the ``done" state and the ``terminated" state of the environment are not distinguished, it will cause critic learning oscillation and result in the performance degradation. Therefore, we set a ``terminated" flag, denoted by $d\in\{0,1\}$, which is a binary variable, to record whether data collection task has been completed by the UAV. In this way, the target value function can be expressed as
	\begin{equation}
	y= r+\left(1-d\right)\cdot \gamma \min_{i=1,2}\left\{ Q_{\theta^\prime_i}\left(s^{\prime},\tilde{a}\right)\right\}.
	\end{equation}
	
	\subsection{Flow of the Proposed Algorithm}\label{flow}
	In Fig. \ref{net_framework}, we show the TD3-TDCTM network framework proposed for the UAV-assisted IoT data collection system. Integrated the {\it information-enhancement}, {\it dimension-spread}, and {\it done-or-terminated} tricks to the TD3 algorithm, we propose a UAV's trajectory optimization algorithm, TD3-TDCTM, to minimize the mission completion time. According to the above discussion, the flow of the TD3-TDCTM algorithm can be described as follow. First, the algorithm randomly initializes the weights $\theta_1$, $\theta_2$, and $\phi$ of the critic and actor networks, respectively ({\it Line} 1) in Algorithm 1. Then, critic and actor target networks are employed to improve the learning stability. The target networks have the same structures as the original actor or critic networks, whose weights $\theta_1^{\prime}$, $\theta_2^{\prime}$, and $\phi^{\prime}$ are initialized in the same manner as their original networks ({\it Line} 2). However, these networks parameters are updated in the entirely different ways. For instance, the target networks' parameters are updated by performing lines 24-25, where a soft target update technique is applied to control the updating rate.
	
	The second part ({\it Lines} 5-13) is the process of exploration. At the beginning of each episode, the algorithm will initialize the environment and receive an initial state and a terminated flag. Then during exploration, the algorithm derives an action from the current actor network $\pi_{\phi}\left(\cdot\right)$ and then add a random noise $\sigma\epsilon$, where $\epsilon$ is a Gaussian noise and $\sigma$ is a decay constant. Through reasonable setting of the exploration noise and the decay factor, we can effectively swing the trade-off between exploration and exploitation. Then, we also need to take care of an important case that an action leads to the boundary violation. To avoid this case, we consider to assign a penalty value $P_{\rm ob}=\frac{1}{K}$ to the pheromone $\zeta_n$ ({\it Lines} 8-10), cancel the corresponding movement (i.e., the UAV stays put without making the movement), and update the corresponding reward and next state accordingly. Besides, we consider another special case that the UAV completes the data collection task in the $n$-th time step, then the rest of the time, denoted by $N_{\rm re}=N_{\max}-n$, is regarded as a positive reward added to the current reward. Finally, we terminate this episode in advance and store the terminated flag as $d=1$, which will be used in the {\it done-or-terminated} technique.
	
	The third part is how to update the neural networks ({\it Lines} 14-26). Similar to those in classical TD3 algorithm, we use a replay buffer, which is initialized at the beginning with the size $R$, for updating the actor and critic networks ({\it Line} 3). Specifically, we first store the collected samples into the replay buffer ({\it Line} 14) and then sample a mini-batch of them from the buffer to update the actor and critic networks ({\it Lines} 16-22). 	\begin{table}[htbp]
		\centering
		\caption{Main simulation parameters.}
		\begin{tabular}{p{.50\columnwidth}p{.30\columnwidth}}
			\toprule  
			\textbf{Simulation parameters}&{\textbf{Value}}\\ 
			\midrule  
			Maximum number of episodes &$M=8000$ \\
			\hline
			The Gaussian distributed exploration noise&$\epsilon\sim \mathcal{N}(0,0.36)$\\
			\hline
			Exploration noise decay rate&$\sigma=0.999$\\
			\hline
			Experience replay buffer capacity &$R=1\times10^5$\\
			\hline
			Target network soft-update rate&$\tau=0.005$\\
			\hline
			Discount factor&$\gamma=0.99$\\
			\hline
			Mini-batch size&$B=256$\\
			\hline
			Actor network and critic networks learning rate&$l_r=0.0001$\\
			\hline
			Maximum time step per episode&$N_{\max}=200$\\
			\bottomrule  
		\end{tabular}
		\label{simulation_tab}
	\end{table}As explained above, the critic networks are updated by minimizing the loss function (Eq. \ref{L}); and the actor network  is updated by computing the gradient (Eq. \ref{J}). 
	
	\section{Simulation Results}\label{IV}
	In this section, numerical results are conducted to evaluate the performance of the proposed TD3-TDCTM algorithm.
	\subsection{Simulation Settings}\label{A}
\begin{figure*}[htbp]
		\centering
		\subfloat[]{
			\label{fig:2d}
			\includegraphics[width=.40\textwidth,keepaspectratio]{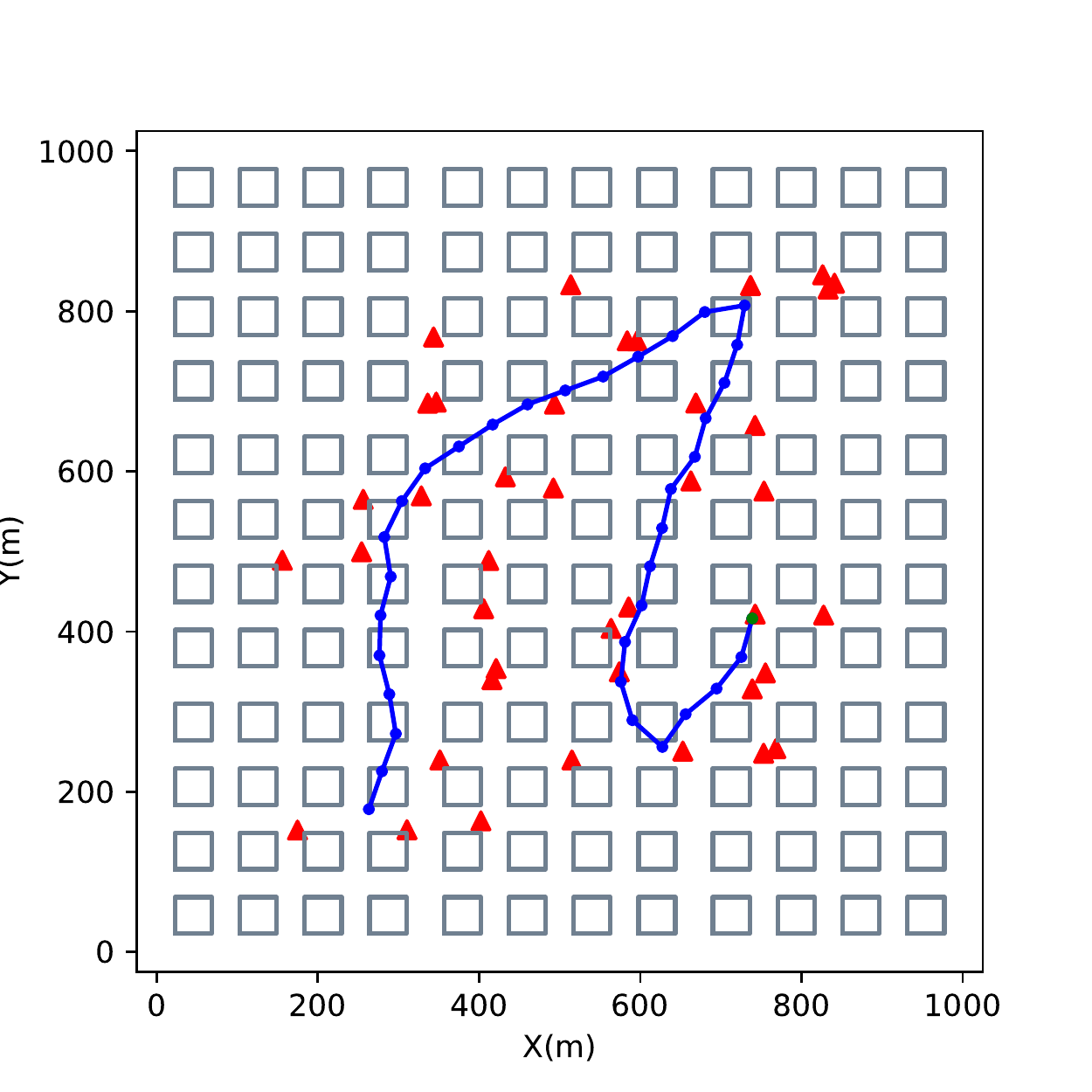}
		}
		\subfloat[]{
			\label{fig:3d}
			\includegraphics[width=.50\textwidth,keepaspectratio]{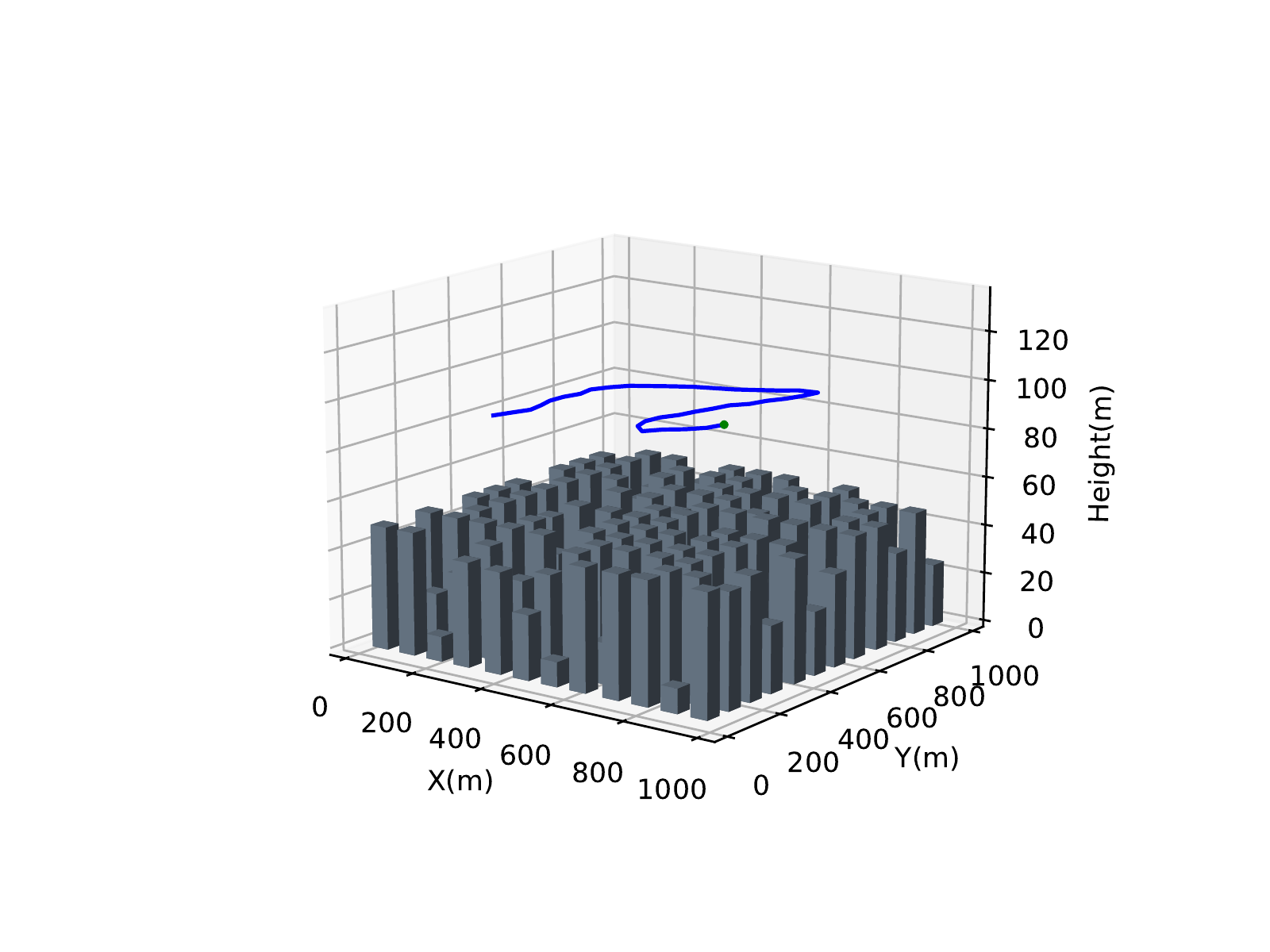}
		}
		\captionsetup{font={footnotesize}, name={Fig.},labelsep=period}
		\caption{UAV's 2D and 3D flight trajectories according to the proposed TD3-TDCTM algorithm, where 40 IoT nodes are considered.}
		\label{fig:view}
	\end{figure*}

	As shown in Fig. \ref{fig:view}, we consider an urban area of size $1,000\times1,000$ $\rm{m}^2$ with the dense and high-rise buildings that are generated by one realization of the statistical model in\cite{ITU}. Figs. 4(a) and 4(b) show the 2D and 3D views of one particular realization of the building locations and heights, respectively, with parameters $\alpha=0.3$, $\beta=144$ buildings/km$^2$, and $\lambda=50$ m. To ensure the practicality, the height of building is clipped to $h\in[10, 50]$ m.
	
	\begin{figure}[htbp]
		\vspace{-6mm}
		\centering{\includegraphics[width=\columnwidth,keepaspectratio]{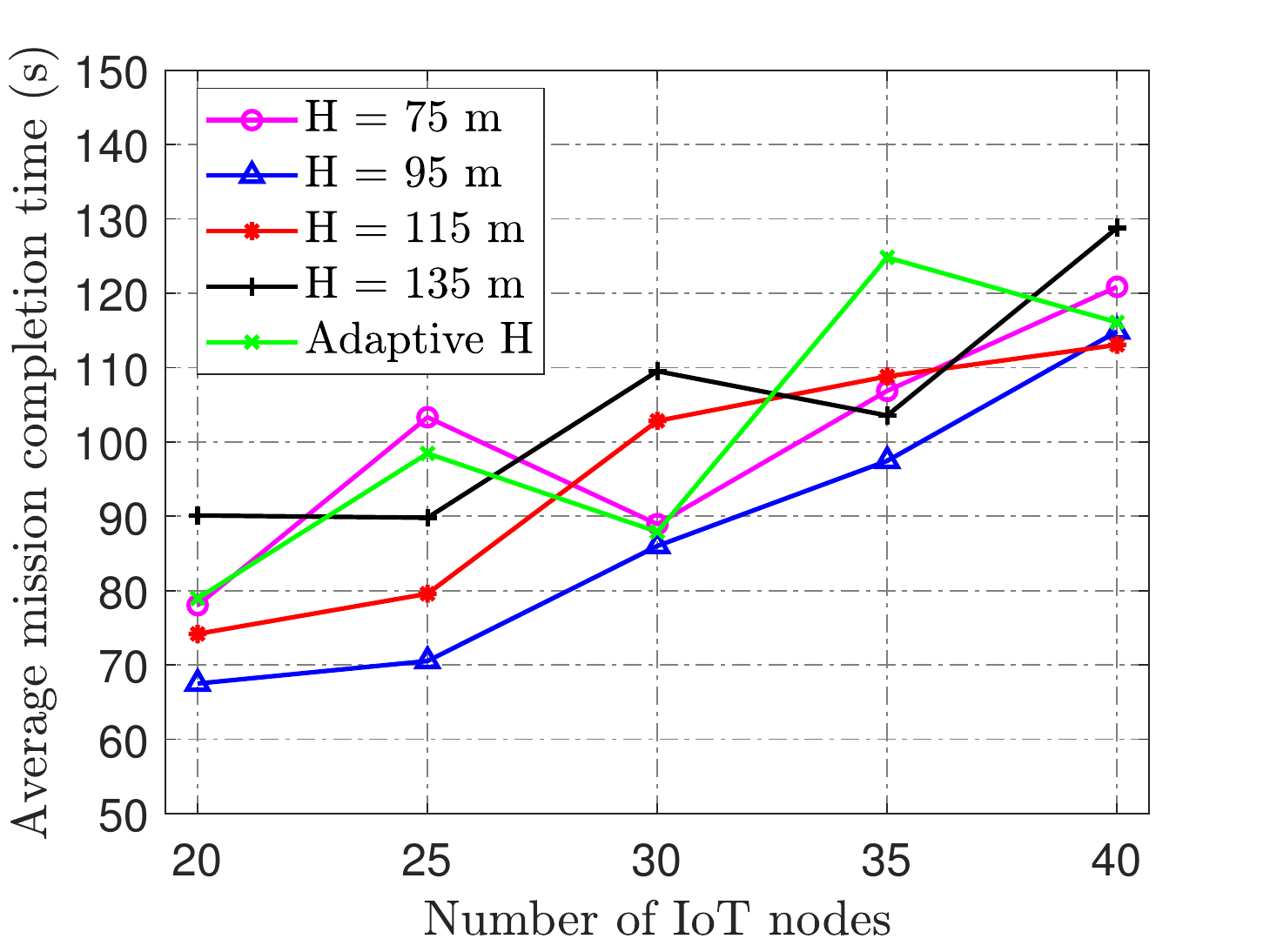}}
		\captionsetup{font={footnotesize}, name={Fig.},labelsep=period}
		\caption{Average mission completion time with different UAV altitudes.}
		\label{fig:DHU}
	\end{figure}
	
	We assume that the transmit power of each IoT node antenna is $P_{\rm Tx}=10$ dBm, the noise power is $P_N=-75$ dBm, the SNR threshold satisfying the basic data collection requirements is $\rho_{\rm th}=0$ dB, the propagation losses are $\eta_{\rm LoS}=0.1$ dB and $\eta_{\rm NLoS}=21$ dB\cite{LAP}, and the information file size of each IoT node is $D_{\rm file}=10$ Mbits. Besides, due to the OFDMA system, the maximal number of IoT nodes that the UAV can serve in the $n$-th time step is $K_{\rm up} = 6$ and the bandwidth allocated to each waken up IoT node is $10$ MHz. The average flight speed of UAV is assumed to be ${\upsilon}\in[0,20]$ m/s, the flight time per step is $\delta_{\rm ft}=2.5 $ s, the hovering time of UAV can be computed by (\ref{eq-ht}), and the fixed altitude of UAV $H$ is a parameter to be optimized. The parameters of pheromone designed are $\kappa_{\rm cov}=10$ and $\kappa_{\rm dis}=1$.

	As for Algorithm 1, all the actor and critic networks are constructed by a 2-layer fully-connected feedforward neural network with 400 neurons. Then the actor network utilizes $\rm \tanh\left(\cdot\right)$ as the activation function to bound the actions. The critic networks utilize the ReLU function for activation. Other major simulation parameters are summarized in Table \ref{simulation_tab} and our simulation runs in the following environment: Python 3.6 and Pytorch 1.4 on a Windows server with 3 NVIDIA 2080TI GPUs.
	
	
	To evaluate the performance, we compare the proposed TD3-TDCTM algorithm with three conventional baseline methods.
	\begin{itemize}
		\item{Scan strategy: The UAV flies according to a preset trajectory which is a rectangular strip track. The track starts from the lower left corner of the area and ends at the upper left corner. Note that such a trajectory design ensures that all locations within the target region are covered by the UAV.}
		
		\item{ACO-based approach: Considering each IoT node as a node, it fixes the initial position of the UAV and exploit the ant colony optimization (ACO) algorithm\cite{ant} to solve the shortest path for completing the routing of each node from the determined starting point.}
		\item{RRT-based approach: Based on the ACO solution, we use a typical search algorithm, named rapidly exploring random tree (RRT) algorithm\cite{RRT} to complete the multi-targets path planning problem. The UAV explores each node in turn according to the ACO sequence. If the data collection of target node and some additional nodes is completed during this period, these nodes will be deleted from the sequence until the sequence is empty. }
	\end{itemize}
	
	\subsection{Result and Analysis}\label{B}
	To verify the effectiveness of our proposed algorithm, we use the trained model for testing. In each simulation realization, the UAV's horizontal position is randomly generated. We totally execute 25 mutually independent realizations, whose outputs are averaged to obtain the final results.
	
	For the UAV-assisted IoT data collection, the establishment of LoS G2A channel is desired for connecting the UAV and IoT node with improved link quality. Typically, the G2A LoS link relies on various factors, such as the density and height of buildings, the locations of both the IoT node and the UAV, as well as the deployment altitude of the UAV. Specifically, the LoS probability is a monotonically increasing function with respect to altitude\cite{LAP}. The higher altitude will enhance the probability of G2A LoS link, but the impact of large-scale fading is more severe. Similarly, the lower altitude will mitigate the impact of large-scale fading, but the probability of G2A LoS link will be reduced obviously. Therefore, we need to make a trade-off between the large-scale fading and the LoS probability so that an optimal altitude can be obtained for maximizing the coverage radius. However, it is hard to obtain an analytical solution for the optimum UAV altitude, which is strongly dependent on the specific urban environment condition. Thus, we first compare the average mission completion time with different UAV altitudes to yield the best altitude in Fig. \ref{fig:DHU}. We can observe that the UAV's altitude $H=95$ m is the best in the considered setting. Besides, we compare the 3D trajectory design scheme of the adaptive adjustment of the flight altitude with that of fixed altitude. Through simulation, we can observe the performance of 3D trajectory is poor compared with the case of fixed flight altitude. This is because the 3D trajectory policy obtained by DRL is still far from the optimal policy. In addition, for buildings of different heights, the optimization of 3D trajectory requires constant adjustment of the UAV altitude during the flight to achieve the optimal coverage. This makes the time consumption on the 3D trajectory of the UAV during the mission is greater than that in the case of fixed flight altitude. Therefore, when serving a large number of distributed IoT nodes, the fixed flight altitude scheme is more suitable for wide area data collection missions. Accordingly, we consider to choose the fixed altitude $H=95$ m in the following simulation.
	
		\begin{figure*}[htbp]
		\centering
		\subfloat[]{
			\label{fig:ACT}
			\includegraphics[width=.5\textwidth,keepaspectratio]{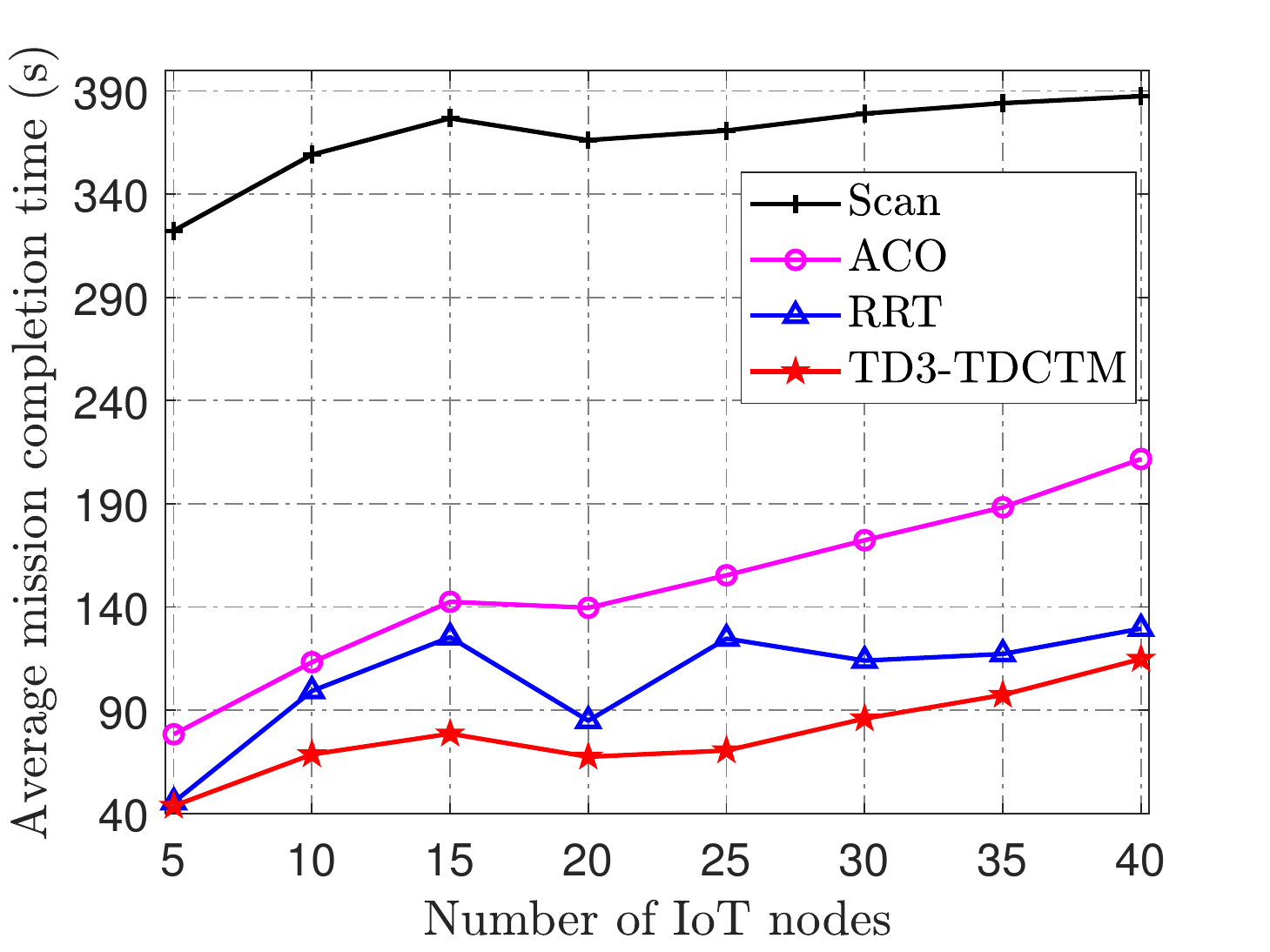}
		}
		\subfloat[]{
			\label{fig:EFC}
			\includegraphics[width=.5\textwidth, keepaspectratio]{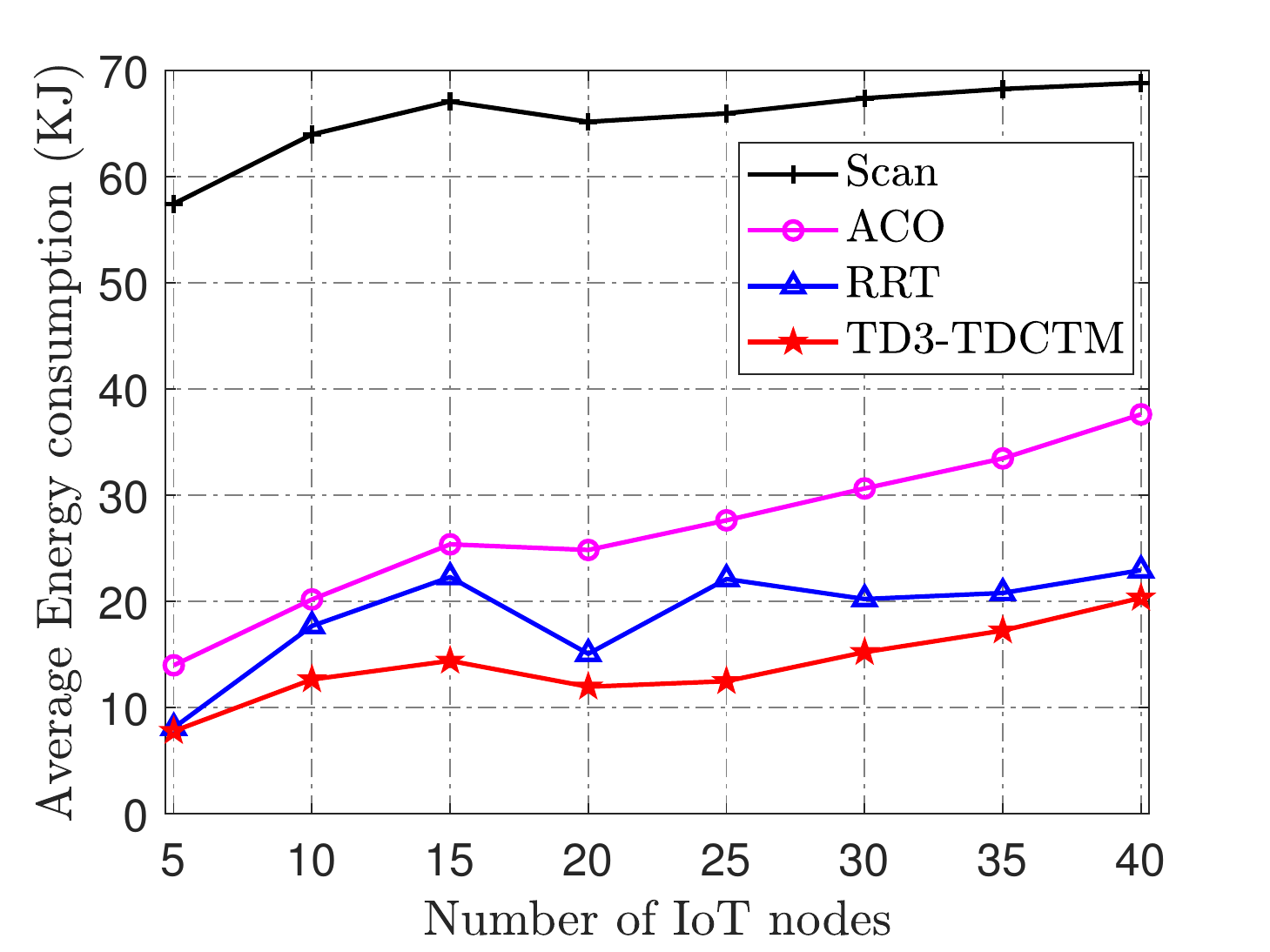}
		}
		\captionsetup{font={footnotesize}, name={Fig.},labelsep=period}
		\caption{The impact of the number of IoT nodes on (a) average mission completion time and (b) average flight energy consumption.}
		\label{fig:perform}
	\end{figure*}
	In Fig. \ref{fig:view}, the UAV's trajectory is plotted under the case of 40 IoT nodes and $H=95$ m, where the red triangles represent the served IoT nodes and the blue curve represents the UAV's trajectory. We can observe that the UAV can complete the data collection mission for all IoT nodes. In such a dense urban environment, buildings are more likely to block the LoS links between the aerial UAV and the terrestrial IoT nodes. Then in the learning process progresses, once the UAV discovers the blockages of LoS links, it would adopt appropriate cruising direction to reestablish the G2A LoS link as soon as possible. This fact shows that TD3-TDCTM algorithm can pilot the UAV to sense and learn the external environment. Therefore, it can learn to obtain an approximately optimal strategy for this practical problem with  imperfect CSI.
	
	In Fig. 6(a), we compare the average mission completion time of different methods versus different numbers of IoT nodes. We can observe that the average mission completion time of proposed TD3-TDCTM algorithm outperforms that of conventional schemes. For 25 IoT nodes, TD3-TDCTM algorithm saves 54.2 s compared with the RRT algorithm, 84.9 s compared with the ACO algorithm, and 300.3 s compared with the Scan strategy. For the Scan strategy, although the UAV can guarantee to serve all IoT nodes, the exceedingly long mission completion time is intolerable. For the ACO algorithm, although it addresses the shortest route problem from the UAV to each IoT node, it does not exploit the sensing ability of the UAV, thus there is still a lot of redundancy in flight trajectory. Developed from the ACO algorithm, the RRT algorithm integrates the exploration mechanism to avoid the unnecessary flight. However, since this exploration is completely random, the reduced mission completion time is not obvious. In contrast, the TD3-TDCTM algorithm can sufficiently and adaptively learn how to adjust the exploration strategy. Therefore, the TD3-TDCTM algorithm can take the minimum time to complete the data collection task, while ensuring each IoT node can be served.
	
	\begin{figure*}[htbp]
		\centering
		\subfloat[]{
			\label{fig:Reward}
			\includegraphics[width=.5\textwidth, keepaspectratio]{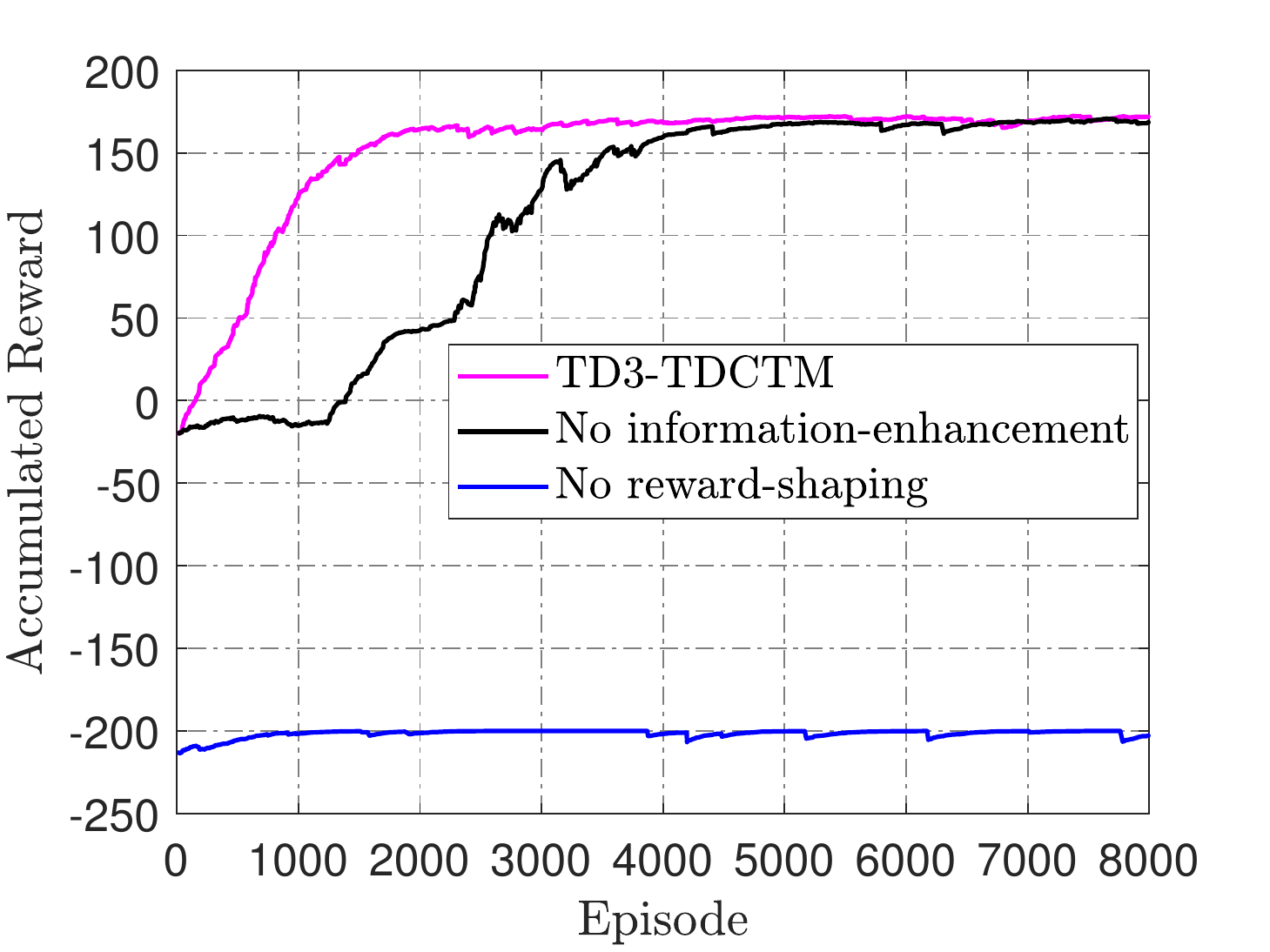}
		}
		\subfloat[]{
			\label{fig:q_loss}
			\includegraphics[width=.5\textwidth,keepaspectratio]{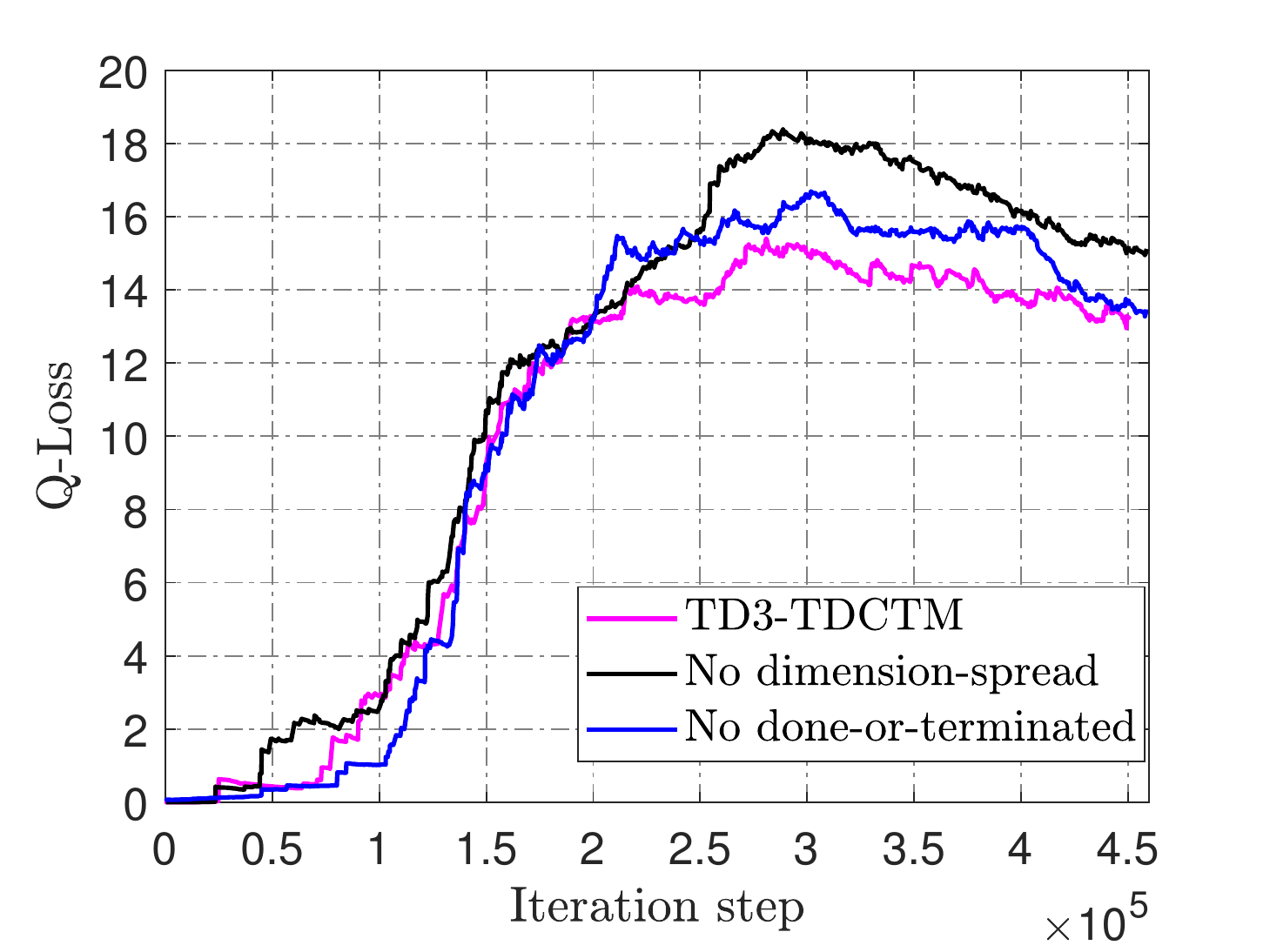}
		}
		\captionsetup{font={footnotesize}, name={Fig.},labelsep=period}
		\caption{Effectiveness of different tricks including {\it information-enhancement}, {\it reward-shaping}, {\it dimension-spread}, and {\it done-or-terminated}, where 40 IoT nodes are considered.}
		\label{fig:technique}
	\end{figure*}

	In order to show the benefit of our proposed approach in terms of energy saving, we compute the energy consumption of the UAV. Specifically, based on \cite{Rate-1}, the propulsion power of the UAV can be modeled as
	\begin{figure}[htbp]
		\centering{\includegraphics[width=\columnwidth,keepaspectratio]{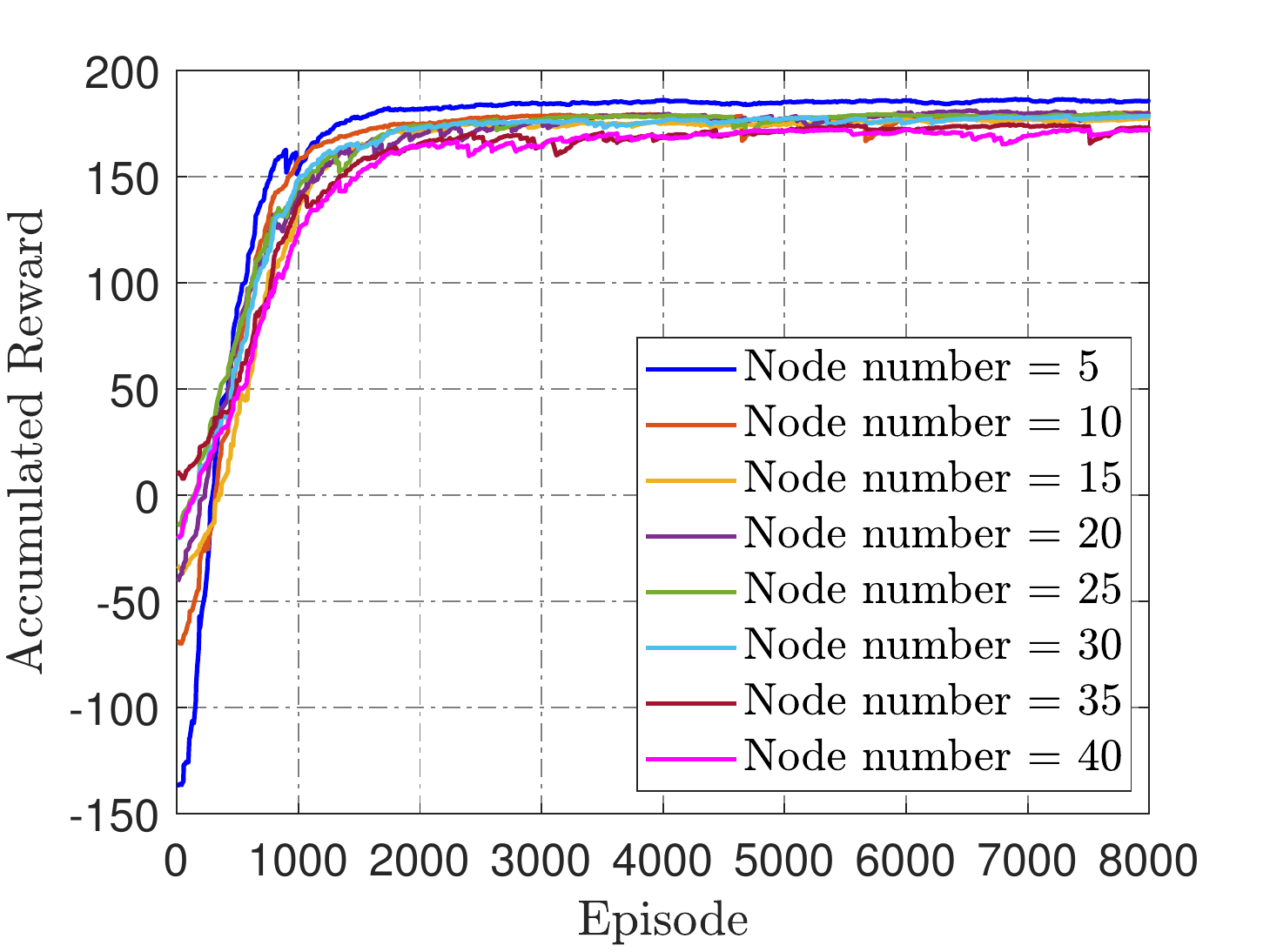}}
		\captionsetup{font={footnotesize}, name={Fig.},labelsep=period}
		\caption{Accumulated reward versus episode with different numbers of IoT nodes.}
		\label{fig:train}
	\end{figure}
	\begin{figure}[htbp]
		\centering{\includegraphics[width=\columnwidth,keepaspectratio]{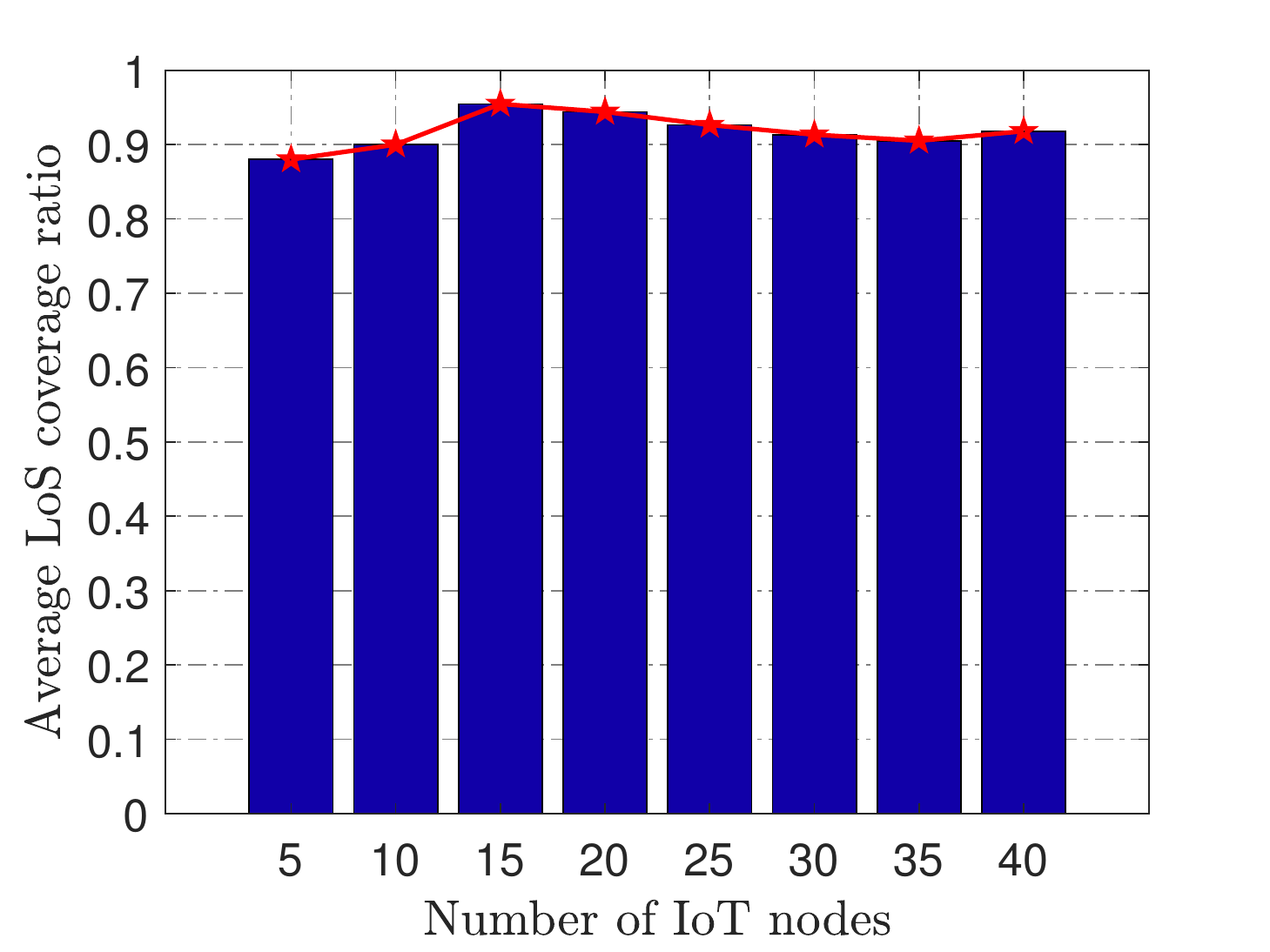}}
		\captionsetup{font={footnotesize}, name={Fig.},labelsep=period}
		\caption{Average LoS coverage ratio with different numbers of IoT nodes.}
		\label{fig:los}
	\end{figure}
	\begin{align}
	P\left(\upsilon\right)=P_0\left(1+\frac{3\upsilon^2}{U_{\rm tip}^2}\right)+&P_1\left(\sqrt{1+\frac{\upsilon^4}{4\upsilon_0^4}}-\frac{\upsilon^2}{2\upsilon_0^2}\right)^{1/2}\notag\\
	&+\frac{1}{2}d_0\varrho sA\upsilon^3,
	\end{align}
	where $\upsilon={\upsilon}_n$ is the horizontal speed of the UAV, $P_0=79.8563$ W and $P_1=88.6279$ W are two constants, $U_{\rm tip}=120$ m/s represents the tip speed of the rotor blade, $\upsilon_0=4.03$ m/s is the mean rotor induced velocity in hover, $d_0=0.6$ and $s=0.05$ are the fuselage drag ratio and rotor solidity, respectively, $\varrho=1.225$  kg/m$^3$ and $A=0.503$ m$^2$ denote the air density and rotor disc area, respectively. Fig. 6(b) shows the average flight energy consumption of different methods versus different numbers of IoT nodes. It is observed that proposed TD3-TDCTM algorithm can save 43.6\% of flight energy compared with the RRT algorithm, 54.9\% compared with the ACO algorithm, and 81.2\% compared with the Scan strategy for 25 IoT nodes. Owing to the substantially reducing the energy consumption of UAV flight, more time and energy can be saved for the UAV to perform data collection.

	In Fig. 7, we demonstrate the effectiveness of the tricks introduced in Section \uppercase\expandafter{\romannumeral3}-D. As shown in Fig. 7(a), we can observe that the TD3-TDCTM algorithm without {\it reward-shaping} does not work well at all in the case of 40 IoT nodes. However, the TD3-TDCTM algorithm with {\it reward-shaping}, like the black curve or the pink curve, can complete the data collection task and converge to a greater and stable reward. It shows that the {\it reward-shaping} mechanism can guarantee the convergence of the proposed algorithm. Besides, without {\it information-enhancement}, the convergence of accumulated reward is obviously slower. Specifically, without {\it information-enhancement}, about 5000 episodes are required to ensure the converge. In contrast, the TD3-TDCTM algorithm only requires about 2000 episodes to reach converge. It shows that the introduction of extra pheromone information as part of the state can enhance the agent's decision efficiency. In Fig. 7(b), we clarify the effectiveness of the other two tricks from the perspective of Q-loss. Specifically, we can see that the Q-loss fluctuation of TD3-TDCTM algorithm without {\it dimension-spread} and without {\it done-or-terminated} is relatively large. Therefore, these two tricks can further improve the convergence performance.
	
	\begin{table}[htbp]
		\centering
		\caption{Impact of different discount factors and buffer sizes on the average completion time.}
		\resizebox{\columnwidth}{!}{
			\begin{tabular}{c|c|c|c|c|c}
				\hline  
				\multirow{2}*{\textbf{Discount factor $\gamma$}}&
				\multicolumn{5}{c}{\textbf{Buffer size $R$}}\\
				\cline{2-6}
				&$0.5\times10^5$&$0.75\times10^5$&$1\times10^5$&$1.25\times10^5$&$1.5\times10^5$\\
				\hline  
				0.8&None&None&None&None&None\\
				\hline
				0.9&79.19&76.72&74.52&74.53&74.33\\
				\hline
				0.99&\textbf{76.45}&\textbf{75.43}&\textbf{71.49}&\textbf{71.29}&\textbf{76.71}\\
				\hline  
		\end{tabular}}
		\label{tab:1}
	\end{table}
	
	\begin{table}[htbp]
		\centering
		\caption{Impact of different neuron numbers and buffer sizes on the average completion time.}
		\resizebox{\columnwidth}{!}{
			\begin{tabular}{c|c|c|c|c|c}
				\hline  %
				\multirow{2}*{\textbf{Neuron number}}&
				\multicolumn{5}{c}{\textbf{Buffer size $R$}}\\
				\cline{2-6}
				&$0.5\times10^5$&$0.75\times10^5$&$1\times10^5$&$1.25\times10^5$&$1.5\times10^5$\\
				\hline  %
				200&76.45&75.43&71.49&71.29&76.71\\
				\hline
				400&73.62&74.1&\textbf{70.52}&73.55&74.96\\
				\hline
				600&85.41&78.4&86.48&77.56&76.36\\
				\hline  %
		\end{tabular}}
		\label{tab:2}
	\end{table}

	Fig. \ref{fig:train} shows the accumulated reward per episode in the training stage under different numbers of IoT nodes. We observe that the accumulated reward shows an upward trend with the increase of the training episodes. After training around 2000 episodes, the accumulated reward gradually becomes smooth and stable. As shown in Fig. \ref{fig:train}, the proposed TD3-TDCTM algorithm has the similar convergence performance in the cases of different numbers of IoT nodes. Hence, the proposed scheme is capable of achieving the good convergence and robustness.
	
	Fig. \ref{fig:los} demonstrates the average LoS coverage ratio of the UAV trajectory designed by the proposed TD3-TDCTM algorithm. We can observe that the probability of completing the data collection tasks using LoS link can always be more than 85\%. It indicates that by using the proposed algorithm, the designed UAV's trajectory can make full use of the environment information so that the G2A LoS link can be established as much as possible. In this way, more IoT nodes can be simultaneously covered along the designed UAV's trajectory.

	In Tables \ref{tab:1} and \ref{tab:2}, we present the experimental results for obtaining the appropriate hyperparameters of the proposed algorithm. Here, we select the discount factor $\gamma$, the number of neurons in each layer in all six networks\footnote{There six networks includes the actor network, target actor network, critic network 1, target critic network 1, critic network 2, and target critic network 2 shown in Fig. \ref{net_framework}.}, and the experience replay buffer size $R$ to be discussed in the case of 25 IoT nodes. As shown in Table \ref{tab:1}, we first present the impact of the discount fact $\gamma$ on the average mission completion time when we set the neuron number to 200. We can observe that the UAV cannot complete the task at all in the case of $\gamma=0.8$. However, when $\gamma$ becomes large, the UAV begins to complete the mission and the average completion time approaches that of the near-optimal solution. This is because a larger $\gamma$ implies a longer-term consideration of future reward. Since our considered scenario is to minimize the mission completion time, increasing $\gamma$ helps the UAV to focus more attention to the possible benefits of future actions and try to navigate in such a way that future movement will complete the current task more quickly.
	
	In Table \ref{tab:2}, we discuss how the neuron number and buffer size affect the mission completion time, under the case of the fixed discount factor $\gamma=0.99$. We can observe that when setting the buffer size as $R=1\times 10^5$, and increasing the neuron numbers from 200 to 400, the average mission completion time will decrease from 71.49 to 70.52, i.e., our model can achieve the optimal performance when neuron number is 400 and buffer size is $R=1\times 10^5$. This is because the appropriate number of neurons can improve the capacity of networks, which can help the agent to better learn the representation of complex correlations among state, action, and reward. Meanwhile, we can conclude that the buffer size $R=1\times 10^5$ can help to achieve a better performance than the rest of setting. On the one hand, the smaller buffer size would lead to an insufficient transition sampling storage or less randomness. On the other hand, the larger buffer size would lead to good samples to be replayed with fewer chances.
	
	\section{Conclusion}
	In this paper, we proposed a DRL based algorithm, TD3-TDCTM, which can design an efficient flight trajectory for a UAV to collect data from IoT nodes in a practical 3D urban environment with imperfect CSI. In particular, we set an additional information, i.e., the merged pheromone, to represent the state information of UAV and environment as a reference of reward which facilitates the algorithm design. By taking the service status of IoT nodes, the UAV's position, and the merged pheromone as input, the TD3-TDCTM algorithm can continuously and adaptively learn how to adjust the UAV's movement strategy for minimizing the completion time under the constraints in flight movement and throughput. Numerical results show a significant performance gain of the TD3-TDCTM algorithm over the existing optimization-based methods. In the future study, we will consider the 3D trajectory design for multi-antenna UAV and multi-UAVs scenarios, where the beamforming and the G2A channel estimation should be considered.

\end{document}